\documentclass[12pt,a4paper,english,superscriptaddress,aps,nofootinbib]{revtex4}
\usepackage[utf8]{inputenc}
\usepackage[T1]{fontenc}
\usepackage{amsmath,amssymb,graphicx}
\makeatletter
\usepackage{babel}
\usepackage[active]{srcltx}
\usepackage{graphicx,color}
\usepackage{changebar}
\usepackage{hyperref}
\usepackage[T1]{fontenc}
\usepackage{esint}
\usepackage{multirow}
\usepackage{dsfont}
\usepackage{ae}
\usepackage{amsmath}
\usepackage{braket}
\usepackage{mathtools}
\usepackage{slashed}
\usepackage{empheq}
\usepackage{multirow}
\usepackage{graphicx}
\usepackage{amsfonts}
\usepackage{amsmath}
\begin{document}
	\title{Geometrically contracted structure in teleparallel $f(T)$ gravity}
	
\author{A. R. P. Moreira}
\email{E-mail: allan.moreira@fisica.ufc.br}
\affiliation{Universidade Federal do Cear\'{a} (UFC), Departamento de F\'{i}sica - Campus do Pici, Fortaleza, CE, C. P. 6030, 60455-760, Brazil.}

 \author{F. C. E. Lima}
\email{E-mail: cleiton.estevao@fisica.ufc.br}
\affiliation{Universidade Federal do Cear\'{a} (UFC), Departamento de F\'{i}sica - Campus do Pici, Fortaleza, CE, C. P. 6030, 60455-760, Brazil.}

\author{C. A. S. Almeida}
\email{E-mail: carlos@fisica.ufc.br}
\affiliation{Universidade Federal do Cear\'{a} (UFC), Departamento de F\'{i}sica - Campus do Pici, Fortaleza, CE, C. P. 6030, 60455-760, Brazil.}

\begin{abstract}
In the teleparallel $f(T)$ gravity scenario, we consider a five-dimensional thick brane. This scenario is interesting because this theory can provide explanations for inflation, radiation and dark matter under certain conditions. It is convenient to assume, for our study, a polynomial profile of the function $f(T)$. Indeed, some polynomial profiles can produce internal structures for which a brane splitting occurs. For functions $f(T)$ with this capability, geometrically contracted matter field configurations are obtained. These contractions of the matter field for the profiles of $f(T)$ reproduce compact-like settings. To complement the study, we analyze the stability of the brane using the concept of Configurational Entropy (CE). The CE arguments are interesting because they tell us the most stable and likely configurations from the brane in this gravitational background. Therefore, we can indicate the best profile of the function $f(T)$.
\end{abstract}

\maketitle

\thispagestyle{empty}

\section{Introduction}

The study of modified gravity models in a braneworld scenario, initially proposed by Randall-Sundrum (RS) \cite{Randall&Sundrum,Randall&Sundrum1}, has attracted much attention \cite{Biggs,Pourhassan,SBS}. This attention is because these models can provide solutions for the hierarchy problem \cite{Randall&Sundrum}, explanations for the cosmological problem \cite{cosmologicalconstant}, and the essence of dark energy and dark matter \cite{darkmatter}. Generally, for these explanations, it is suggested that the thick brane is coupled to the scalar field \cite{Goldberger1999,Gremm1999, DeWolfe1999,Dzhunushaliev2009,Charmousis2001, Arias2002, Barcelo2003, CastilloFelisola2004, Navarro2004, BarbosaCendejas2005, Koerber2008, Chinaglia2017}. It is interesting to note that all these works consider only the contribution of the spacetime curvature without torsion. Otherwise, it is necessary to use a teleparallel equivalent of general relativity (TEGR) \cite{Hayashi1979, deAndrade1997, deAndrade1999, Aldrovandi}. 

A particularly interesting class of modified gravity theory is the teleparallel theory \cite{Hayashi,Hehl,Kop}. Among the variety of teleparallel models, we have the $f(T)$ theory. In $f(T)$ theory, it is conventional to use the tetrad field to define the Weitzenb\"{o}ck connection. The Weitzenb\"{o}ck connection represents torsion without curvature. This characteristic allows a proposal to emerge to justify the accelerated expansion of the universe \cite{Ferraro2007, Ferraro2011}. The $f(T)$ gravity has presented significant results in several scenarios, such as cosmology \cite{ft}, dark energy \cite{ftdarkenergy1, ftdarkenergy2}, gravitational waves \cite{ftgw,ftgw2,ftgw3}, black holes \cite{Miao,blackhole1,blackhole2,Nashed2021}, and braneworld \cite{Yang2012, ftnoncanonicalscalar,ftborninfeld,ftenergyconditions,tensorperturbations, Moreira2021a, Moreira2022a}. In particular, in the braneworld scenario, it was possible to observe the formation of internal structures with only a single scalar field as a source \cite{MLA11}. In general relativity, the emergence of these structures is possible with at least two interacting scalar fields. This result, in particular, motivates us to analyze the conditions that lead to the geometrically contracted structures in a braneworld scenario in a $f(T)$ gravity. To help us in this analysis, we use Differential Configurational Entropy (DCE) to obtain the balance and stability configurations of the model.

Differential Configurational Entropy (DCE) is one of the variants of Configurational Entropy (CE). Indeed, the arguments from the CE have been shown to be useful for studying the stability of field configurations in braneworld \cite{MLA11}. Furthermore, CE applications are good tools for the physical understanding of some systems, see e. g., Refs. \cite{COLee,COLee1}. The interest in the CE is because it can provide information on the parameters of a given model to build a stable field configuration \cite{GleiserSowinki,GS,Correa2015a}. In Ref. \cite{GS}, the authors show that the higher (or lower) the CE value, the higher (or lower) the energy value that approximates the real solution. The CE reported significant results that helped us understand the dynamics of spontaneous symmetry breaking \cite {G1}. Moreover, CE is applied to compact objects to analyze the stability limits \cite{G2,G3}. The stability of modified gravity models in braneworld scenarios was also analyzed using the DCE, see Refs. \cite{Correa2015b,Correa2015c,Correa2016b}.

The paper is organized as follows: In Sec. \ref{sec1}, we discussed the concept of $f(T)$ teleparallel gravity in the braneworld scenario. In addition, the energy densities and matter field solutions are inspected. In Sec. \ref{sec2}, it is made a brief review of the concepts of CE and DCE. Next, we study the DCE of the models displayed. Finally, in Sec. \ref{sec3}, we discussed our findings.

\section{Teleparallel $f(T)$ gravity: Braneworld}
\label{sec1}

To study the brane model in $f(T)$ gravity is necessary to review some concepts of teleparallel gravity. In teleparallelism, the dynamic variable is {\it vielbein} or tetrad fields, unlike General Relativity (GR), where the dynamic variable is the metric. However, we can relate the {\it vielbein} to the metric as follows \cite{Aldrovandi}
\begin{align}
g_{MN}=\eta_{ab}h^a\ _M h^b\ _N.
\end{align}
The {\it vielbein} is an orthonormal base in the tangent space. The uppercase Latin letter ($M=0,...,D-1$) represents the indices of the bulk coordinates. On the other hand, the lowercase Latin indices ($a=0,...,D-1$) represent the indices of the tangent space coordinates.

Using the \textit{vielbein}, we can build the Weitzenb\"{o}ck connection, namely, 
\begin{align}
\widetilde{\Gamma}^P\ _{NM}=h_a\ ^P\partial_M h^a\ _N,
\end{align}
which is an relevant connection for teleparallelism \cite{Aldrovandi}. By Weitzenb\"{o}ck connection, let us build the torsion tensor as
\begin{align}\label{tt}
T^{P}\  _{MN}= \widetilde{\Gamma}^P\ _{NM}-\widetilde{\Gamma}^P\ _{MN}.
\end{align}

Adopting the torsion tensor (\ref{tt}), the contorsion tensor \cite{Aldrovandi} as defined as
\begin{align}
K^P\ _{NM}=\frac{1}{2}\Big( T_N\ ^P\ _M +T_M\ ^P\ _N - T^P\ _{NM}\Big),
\end{align}
and the dual torsion tensor \cite{Aldrovandi} is
\begin{align}
S_{P}\ ^{MN}=\frac{1}{2}\Big( K^{MN}\ _{P}-\delta^N_P T^{QM}\ _Q+\delta^M_P T^{QN}\ _Q\Big).
\end{align} 
In GR, the theory connection is the Levi-Civita connection ($\Gamma^P\ _{NM}$). It is worth mentioning that the Levi-Civita connection is related to the Weitzenböck connection by the contortion tensor, namely,  $\Gamma^P\ _{NM}=\widetilde{\Gamma}^P\ _{NM}-K^P\ _{NM}$ \cite{Aldrovandi}.

In teleparallel gravity, the Lagrangian is
\begin{align}
\mathcal{L}=-h\frac{T}{4\kappa_g},
\end{align} 
where $T=T_{PMN}S^{PMN}$ is the torsion scalar, $\kappa_g=4\pi G/c^4$ is the gravitational constant, and $h=\sqrt{-g}$.

After all these definitions, we can then construct the Lagrangian of the $f(T)$ gravity, i. e., $\mathcal{L}=-hf(T)/4\kappa_g $.% \cite{Moreira2021aa}. Therefore, the gravitational action is 
\begin{align}\label{55.5}
\mathcal{S}=-\frac{1}{4\kappa_g}\int h \Big[f(T)+4\kappa_g \mathcal{L}_m\Big]d^5x,
\end{align}
where $\mathcal{L}_m$ is the matter Lagrangian. Using the action (\ref{55.5}) is obtained the modified gravitational field equation \cite{Yang2012}, namely,
\begin{align}\label{3.36}
\frac{1}{h}f_T\Big[\partial_Q\left(h S_N\ ^{MQ}\right)-h\widetilde{\Gamma}^R\ _{SN}S_R\ ^{MS}\Big]-f_{TT}S_N\ ^{MQ}\partial_Q T+\frac{1}{4}\delta_N^Mf=-\kappa_g\mathcal{T}_{N}\,^{M},
\end{align}
where $f\equiv f(T)$, $f_T\equiv\partial f(T)/\partial T$, $f_{TT}\equiv\partial^2 f(T)/\partial T^2$, and $\mathcal{T}_N\ ^M$ is the stress-energy tensor.

Our purpose is to analyze the braneworld scenario. Thinking about this, let us propose the ans\"{a}tz for the metric as follows
\begin{align}\label{45.a}
ds^2=e^{2A(y)}\eta_{\mu\nu}dx^\mu dx^\nu+dy^2.
\end{align}
Here,  $e^{A(y)}$ is the warp factor. This factor is responsible for controlling the brane width, and $\eta_{\mu\nu}=(-1, 1, 1, 1)$ is the usual metric of Minkowski spacetime.

The dynamic variable which interests us is the {\it veilbein}. In this case, adopting the metric (\ref{45.a}) the {\it vielbein} will be
\begin{align}\label{45.aaa}
h^a\ _M = \text{diag} (e^A,e^A,e^A,e^A, 1).
\end{align}
This {\it veilbein} represents a good choice between all possibilities {\it veilbein}. In fact, the {\it veilbein} (\ref{45.aaa}), was used in Refs.\cite{Yang2012,ftnoncanonicalscalar,ftborninfeld, ftenergyconditions, tensorperturbations,ftmimetic}, because with it, the equations of the gravitational field do not present additional restrictions to the functions $f(T)$ nor to the torsion scalar.

Allow us to use a matter Lagrangian described by a single real scalar field, i. e., 
\begin{align}
\mathcal{L}_m=\frac{1}{2}\partial^M\phi\partial_M\phi+V(\phi),
\end{align}
with $\phi\equiv \phi(y)$.

Now, let us propose the following profiles for the function $f(T)$:
\begin{align}\label{f1}
f_1(T)=&T+kT^{n},\\ \label{f2}
f_2(T)=&T+\alpha T^{2}+\beta T^{3}.
\end{align}
These models are the simplest and represent a generalization of teleparallel gravity. Note that the models presented in Eqs. (\ref{f1}) and (\ref{f2}) are prototypes that allow a modification of the usual teleparallel theory. This change occurs by adjusting the $k$, $n$, $\alpha$, and $\beta$ parameters. To simplify our analysis, we consider, without prejudice, the gravitational constant $\kappa_g=1$.

For the first model $f_1(T)$, the gravitational field equations are:
\begin{align}\label{w.0}
&\phi''+4A'\phi=\frac{d V}{d\phi},\\
\label{w.1}
&\Big(1+B_nknA'^{2(n-1)}\Big)A''= -\frac{2}{3}\phi'^2,\\
\label{w.2}
&\Big(1+B_nkA'^{2(n-1)}\Big)A'^2= \frac{1}{3}\Big(\frac{1}{2}\phi'^2-V\Big),
\end{align} 
where $B_n=(-1)^{n-1}12^{n-1}(2n-1)$. Here, the prime notation $ (\ '\ ) $ represents derivative with respect to variable $y$.

Furthermore, one notes that when $k=0$, we obtain the equations of the usual teleparallelism, i. e.,
\begin{align}
\label{ww.1}
A''= -\frac{2}{3}\phi'^2,
\end{align}
and
\begin{align}
\label{ww.2}
A'^2= \frac{1}{3}\Big(\frac{1}{2}\phi'^2-V\Big).
\end{align}
Indeed, these expressions are the equivalent equations of the GR ( see Refs.
\cite{DeWolfe1999,Csaki2000aa,Csaki2000ab}).

The energy density is defined as 
\begin{align}\label{3333}
\rho(y)=-e^{2A}\mathcal{L}_m.
\end{align}

Utilizing the Eqs. (\ref{w.1}) and (\ref{w.2}), we have
\begin{align}\label{323454}
\rho(y)=-\frac{3}{2}\frac{d}{dy}\Big[A'\Big(1+B_nknA'^{2(n-1)}\Big)e^{2A}\Big]-3A'^{2n}\Big[\frac{B_n(n-1)k}{2n-1}\Big]e^{2A}.
\end{align}
Perceive that if $k=0$, the result of the GR is obtained, so that the Eq. (\ref{323454}) it will be a total derivative \cite{DeWolfe1999,Csaki2000aa,Csaki2000ab}. However, everything changes when $k\neq 0$. If $k\neq 0$, the second term contributes to the energy, so the energy is 
\begin{align}\label{324}
E=-3\Big[\frac{B_n(n-1)k}{2n-1}\Big]\int e^{2A}A'^{2n}dy.
\end{align}

For the second model $f_2(T)$, the gravitational field equations are
\begin{align}\label{w.220}
&\phi''+4A'\phi=\frac{d V}{d\phi},\\
\label{w.221}
&\Big[1-72A'^{2}(\alpha-30\beta A'^{2})\Big]A''= -\frac{2}{3}\phi'^2,\\
\label{w.222}
&\Big[1+-36A'^{2}(\alpha-20\beta A'^{2})\Big]A'^2= \frac{1}{3}\Big(\frac{1}{2}\phi'^2-V\Big).
\end{align} 
Note that if $\alpha=\beta=0$, the equivalent equations of GR [Eqs. (\ref{ww.1}) and (\ref{ww.2})] are obtained.

Our purpose is to study the geometrically contracted strutures (and its phase transition) in a thick brane. To achieve this goal, allow us to propose two types of warp factor, namely, 
\begin{align}
\label{coreA}
A_1(y)&=-p\ln{\cosh(\lambda y)},\\
A_2(y)&=\ln\vert\tanh[\lambda(y+c)]-\tanh[\lambda(y-c)]\vert.
\end{align}
The ans\"{a}tz $A_1(y)$ was previously used by Guo et al. \cite{ftmimetic} in the study of thick branes in $f(T)$ mimetic gravity. The parameter $\lambda$ has a mass dimension, and $p$ is an integer parameter. On the other hand, Tan et al. \cite{Tan} used the ans\"{a}tz $A_2(y)$ for the study of gravitational resonances on $f(T)$-branes. In the case of the warp factor $A_2(y)$, the parameter $c$ is a unit of length and represents the distance of two sub-branes. In both cases, the ans\"{a}tze depicts an asymptotically AdS$_5$ spacetime.

Before starting the study of the particular cases, allow us to highlight that verifying the results of the $f(T)$ theory exposed here can be performed by cosmographic observations \cite{Capozziello:2019cav, Capozziello:2017uam}. That is possible because the cosmological principle requires a scale factor as the only degree of freedom that governs the universe \cite{Capozziello:2019cav}. Thus, one can expand this factor (namely of $a(t)$) as
\begin{align}\label{Afactor}
a(t)=1+\sum_{k=1}^{\infty}\frac{1}{k!}\frac{d^ka}{dt^k}\bigg\vert_{t=t_0}(t-t_0)^k.
\end{align}
Using $a(t)$, one defines, respectively, the Hubble, deceleration,  jerk, and snap parameters as
\begin{align}
H(t)=\frac{1}{a}\frac{da}{dt}, \hspace{0.5cm} q(t)=-\frac{1}{aH^2}\frac{d^2a}{dt^2}, \hspace{0.5cm} j(t)=\frac{1}{aH^3}\frac{d^3a}{dt^3}, \hspace{0.5cm} \text{and} \hspace{0.5cm} s(t)=\frac{1}{aH^4}\frac{d^4a}{dt^4}.
\end{align}
As seen in Refs. \cite{Capozziello:2019cav, Capozziello:2017uam} , one can describe the physical properties of these coefficients using the form of Hubble's expansion \cite{Capozziello:2019cav}. Furthermore, torsion is known to be related to Hubble's factor, i. e., $T\propto H^2$. Thus, we can use this relation and its inverse to obtain a specific form of the function $f(T)$. Or yet, apply the inverse steps searching to obtain the profile of the cosmological parameters. So, using this approach, it is possible to confront the $f(T)$ gravity theories here exposed with cosmography data. For more details, see Refs. \cite{Capozziello:2019cav, Capozziello:2017uam} .

\subsection{The case $A_1(y)=-p\ln{\cosh(\lambda y)}$}

For the $f_1(T)$ model, the Eqs. (\ref{w.1}) and (\ref{w.2}) are rewritten as
\begin{align}\label{q.5}
&\phi'^2(y)=\frac{3}{2}p\lambda^2\mathrm{sech}^2(\lambda y)\Big\{1+(-1)^{2(n-1)}B_nkn\Big[p\lambda\tanh(\lambda y)\Big]^{2(n-1)}\Big\},
\end{align}
\begin{align}
\label{q.55} \nonumber
V(\phi(y))=&\frac{3}{4}p\lambda^2\mathrm{sech}^2(\lambda y)-3[p\lambda\tanh(\lambda y)]^2+\frac{(-1)^{2n}3B_nk}{4p}
\Bigg\{\Big[4p-n\ \mathrm{csch}^2(\lambda y)\Big]\times \\ &\Big[p\lambda\tanh(\lambda y)\Big]^{2n}\Bigg\}.
\end{align}

Through the equation (\ref{3333}), the energy density is
\begin{align}\label{121222}\nonumber
\rho(y)=&\frac{3}{2p}\Bigg\{(p\lambda)^2\Big[2p\tanh^2(\lambda y)-\mathrm{sech}^2(\lambda y)\Big]+(-1)^{2n}B_nk\Big[2p-n\ \mathrm{csch}^2(\lambda y)\Big]\times\\
&\Big[p\lambda\tanh(\lambda y)\Big]^{2n}\Bigg\}\cosh^{-2p}(\lambda y).
\end{align}

\begin{figure}[ht!]
\begin{center}
\begin{tabular}{ccc}
\includegraphics[height=4cm,width=5.2cm]{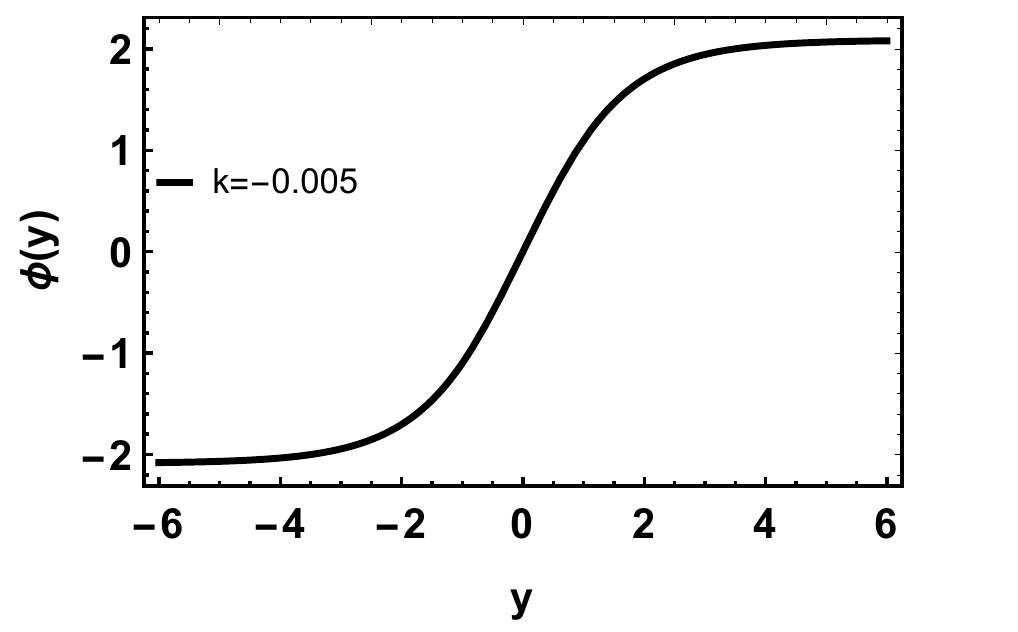}
\includegraphics[height=4cm,width=5.2cm]{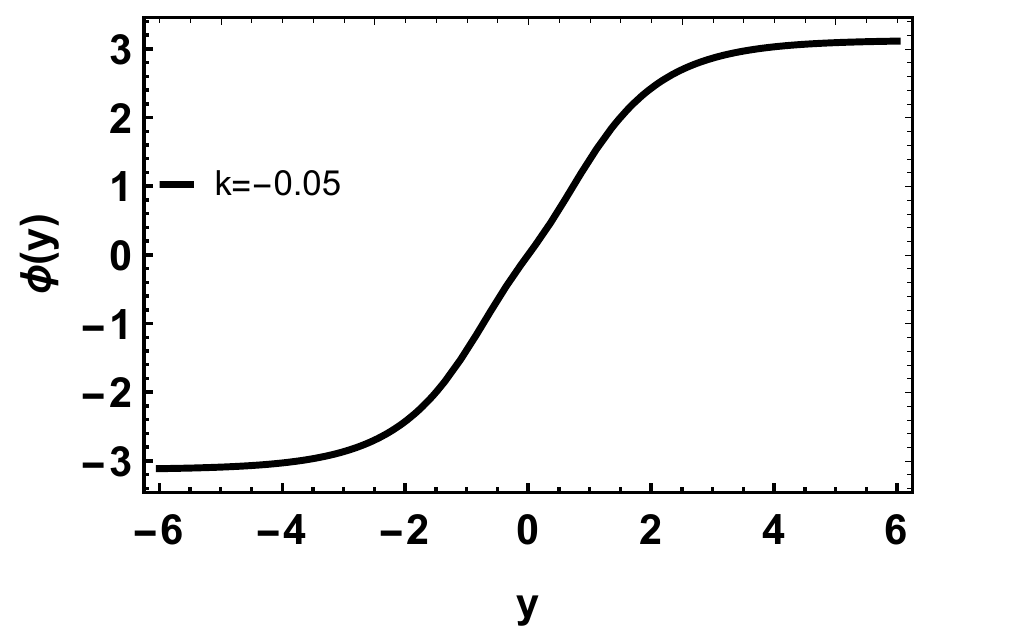}
\includegraphics[height=4cm,width=5.2cm]{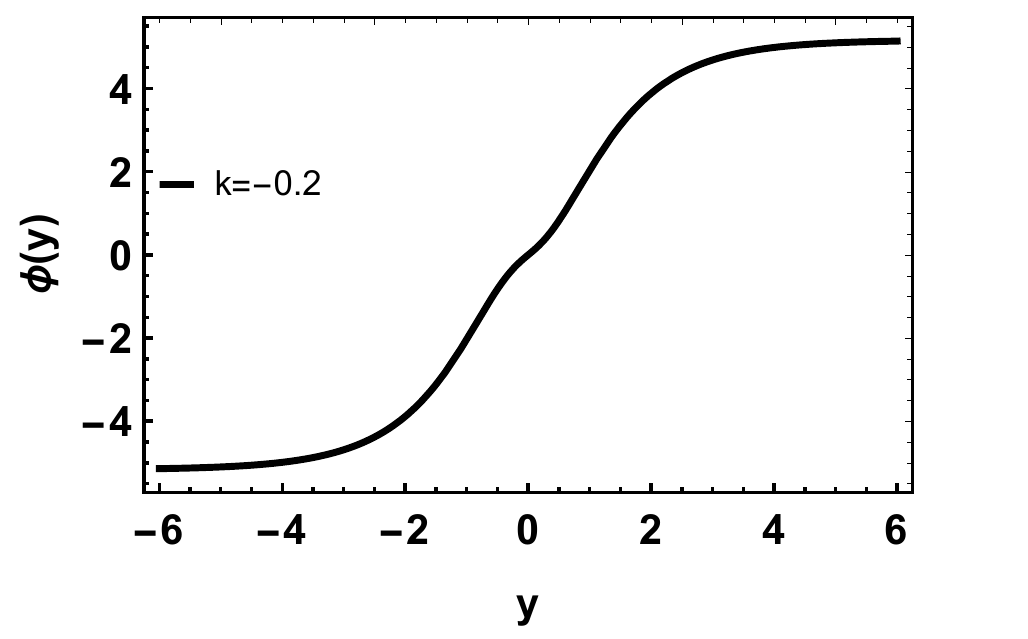}
\end{tabular}
\end{center}
\vspace{-0.5cm}
\caption{
Scalar field solution for $n=2$ with $p=\lambda=1$.
\label{fig1}}
\end{figure}

\begin{figure}[ht!]
\begin{center}
\begin{tabular}{ccc}
\includegraphics[height=5.5cm,width=7cm]{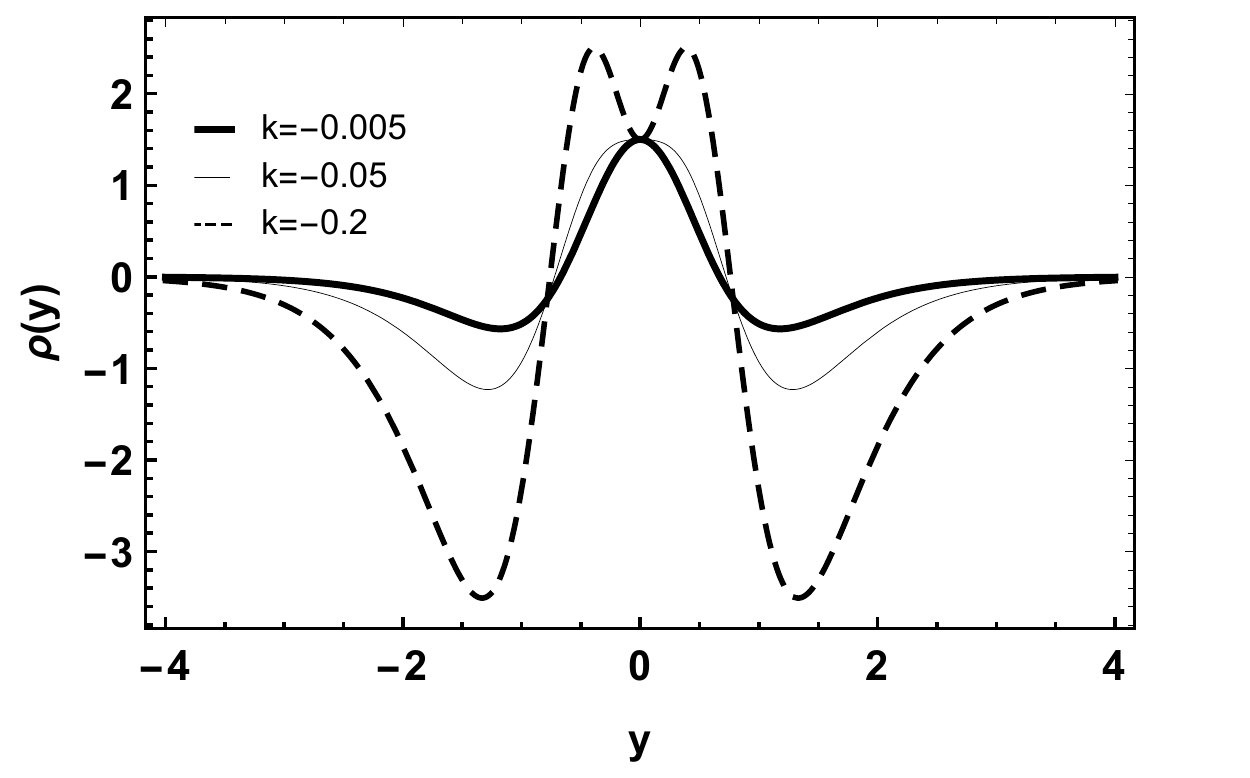}
\includegraphics[height=5.5cm,width=7cm]{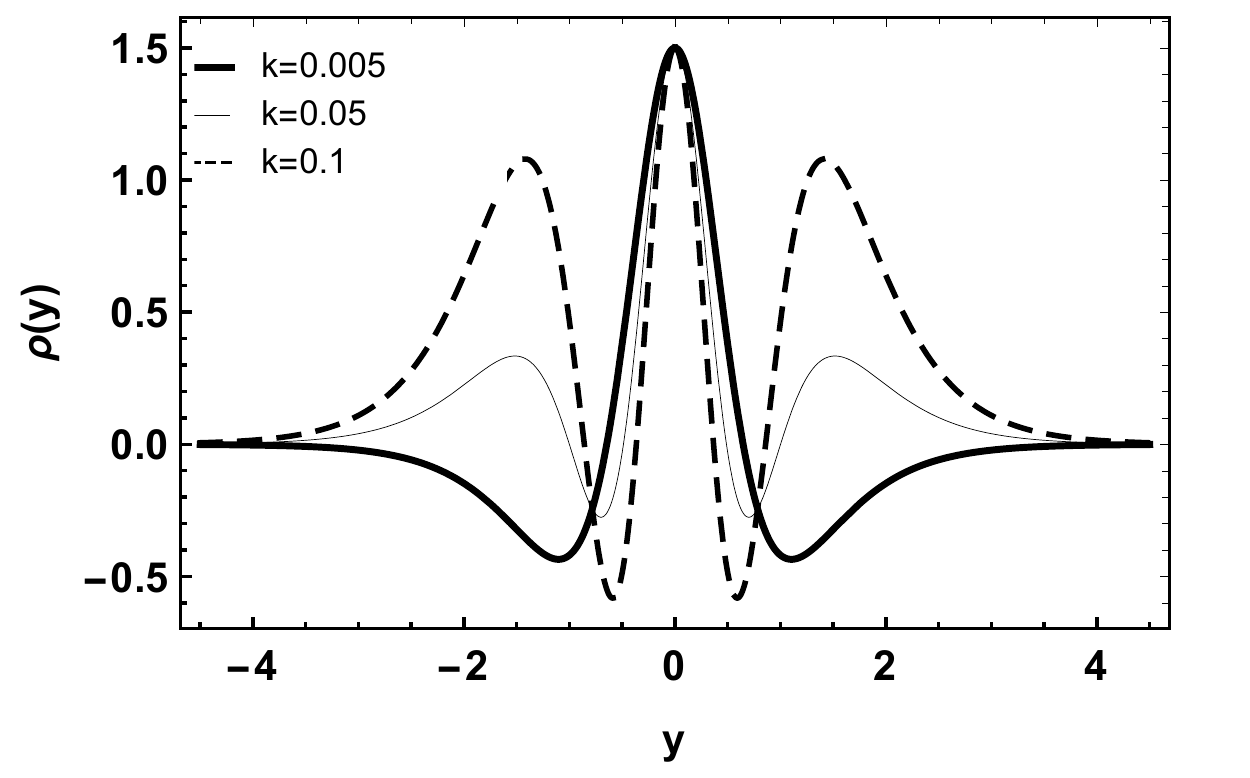}\\ 
(a) \hspace{7cm} (b)
\end{tabular}
\end{center}
\vspace{-0.5cm}
\caption{Brane energy density for $n=2$ with $p=\lambda=1$. (a) The case $k<0$. (b) The case $k>0$.
\label{fig2}}
\end{figure}

To find the solution of the scalar field, we need to solve the equation (\ref{q.5}). The analytical solutions of the scalar field are complicated to obtain. For simplicity, for our analysis, we numerically calculate. In Fig. \ref{fig1} is displayed the behavior of the scalar field for $n=2$. In turn, the behavior of energy density is shown in Fig. \ref{fig2} for $n=2$. 

Looking at the behavior of the scalar field (Fig. \ref{fig1}), one notices that if $k<0$, the solution of the scalar field is continuously transforming from a kink-like structure to a double-kink-like. The interpretation of this behavior is that the matter field will suffer some phase transitions. The emergence of double-kink structures leads us to the hypothesis of multiple phase transitions, and consequently, a geometric contraction of the kink. We will discuss this hypothesis later in Sec. \ref{sec2}. This behavior of the matter field is felt (or detected) in the energy density since the energy density starts to have two critical points (peaks) of energy (Fig. \ref{fig2}(a)). In summary, each peak that appears in the energy density corresponds to a kink-like solution. It is important to note that these energy peaks also represent brane splitting. Finally, if $k\approx -0.05$, it is noticed that double-kink-like solutions begin to appear. Meanwhile, for $k>0$, the appearance of internal structures is observed (Fig. \ref{fig2}(b)) when $k\approx 0.05$.

\begin{figure}[ht!]
\begin{center}
\begin{tabular}{ccc}
\includegraphics[height=4cm,width=5.2cm]{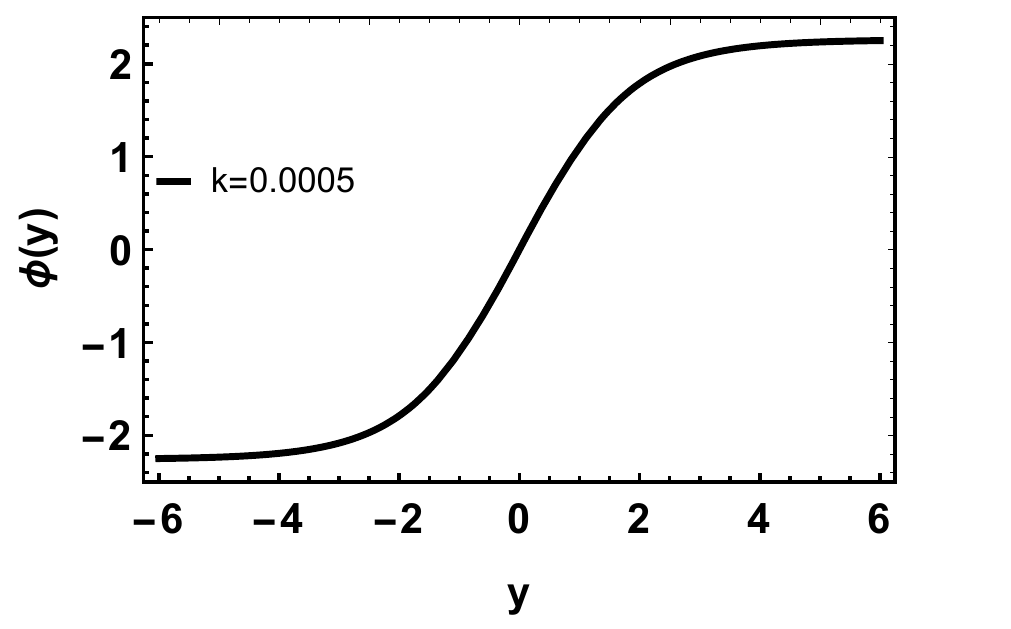}
\includegraphics[height=4cm,width=5.2cm]{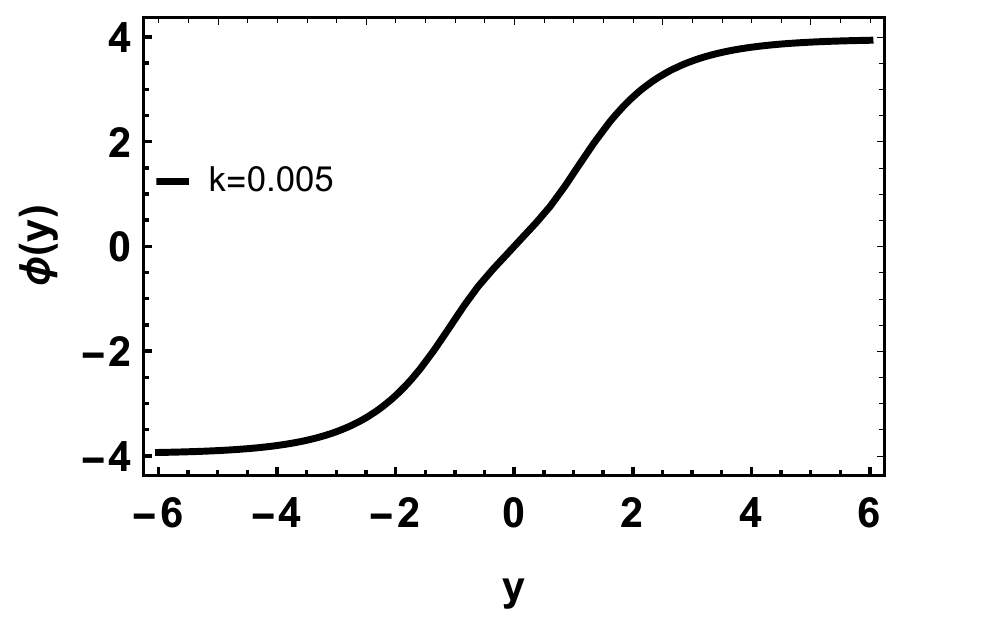}
\includegraphics[height=4cm,width=5.2cm]{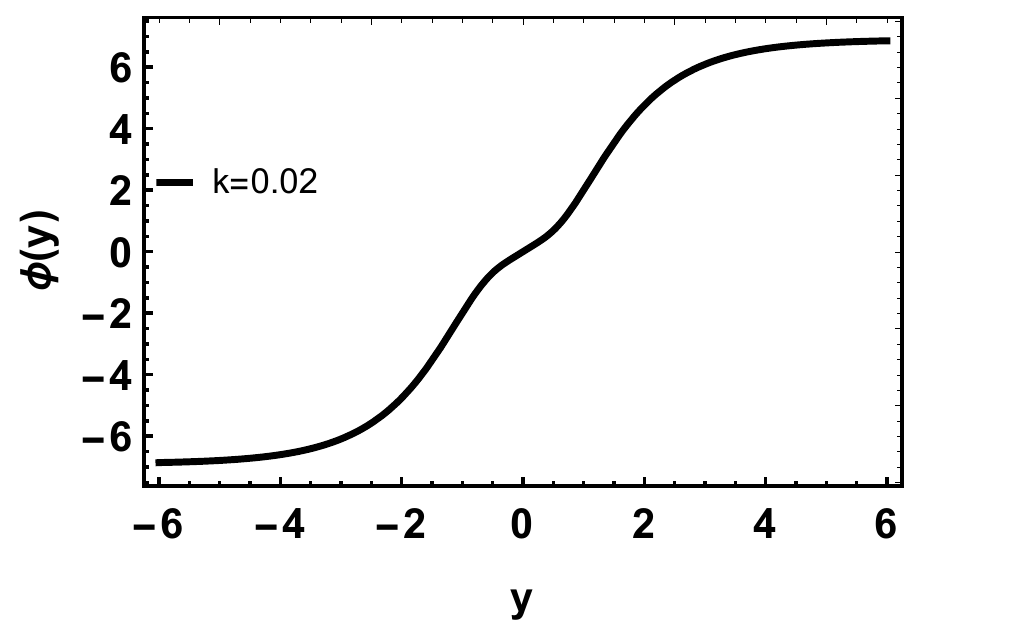}
\end{tabular}
\end{center}
\vspace{-0.5cm}
\caption{Scalar field solution for $n=3$ with $p=\lambda=1$.
\label{fig3}}
\end{figure}

\begin{figure}[ht!]
\begin{center}
\begin{tabular}{ccc}
\includegraphics[height=5.5cm,width=7cm]{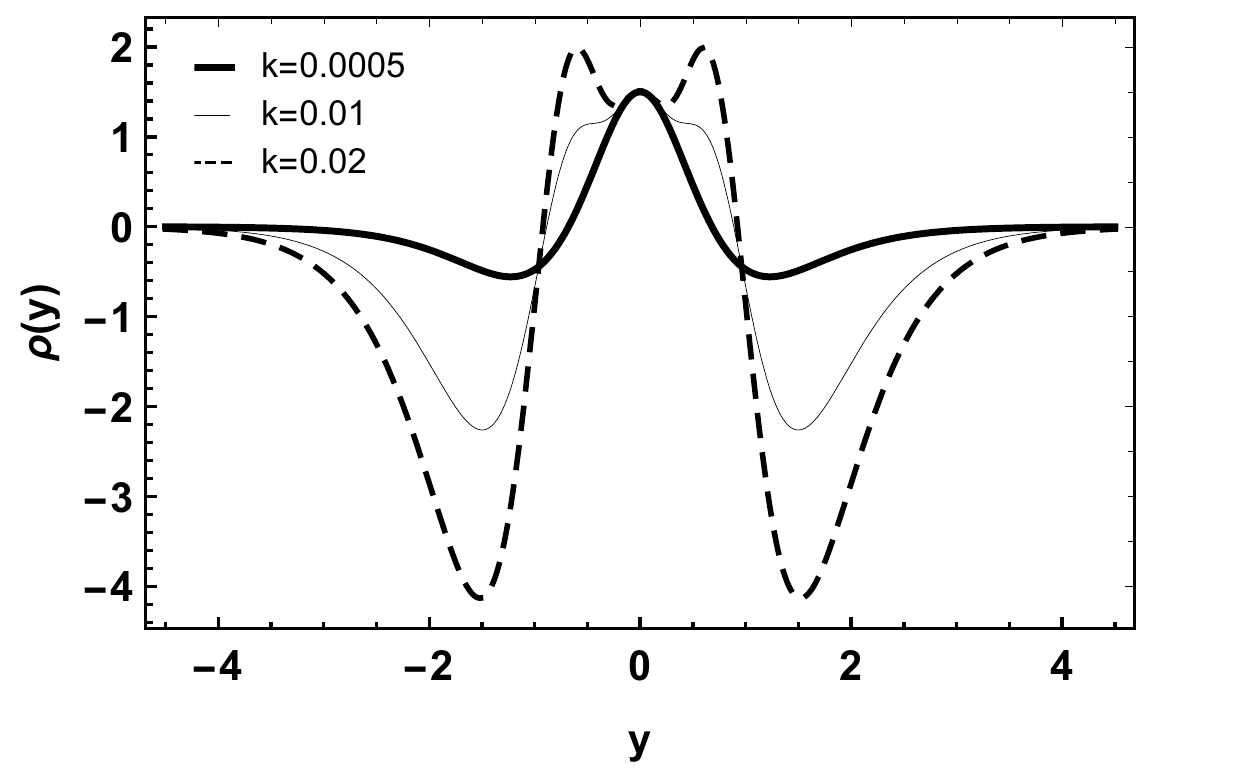}
\includegraphics[height=5.5cm,width=7cm]{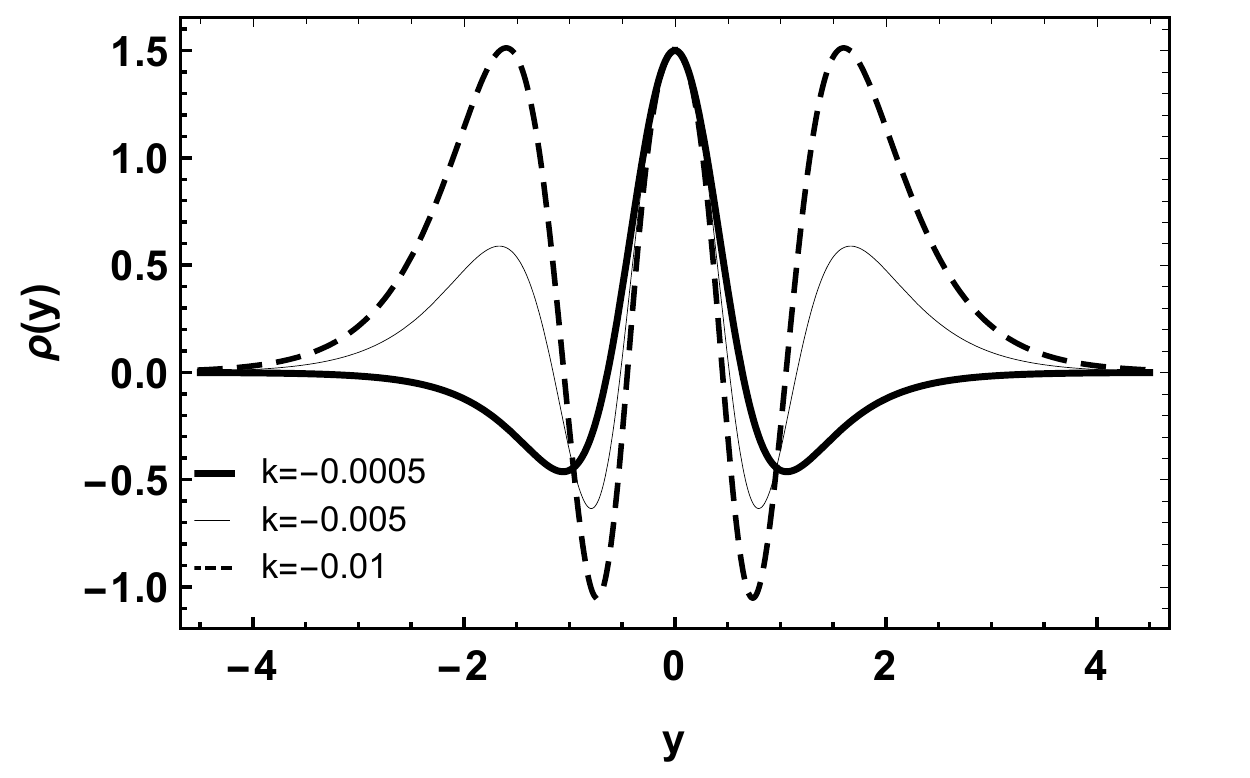}\\ 
(a) \hspace{7cm} (b)
\end{tabular}
\end{center}
\vspace{-0.5cm}
\caption{Brane energy density for $n=3$ with $p=\lambda=1$. (a) The case $k>0$. (b) The case $k<0$.
\label{fig4}}
\end{figure}

For $n=3$ and $k>0$, we perceived that with a decrease of the value of $k$, the profile of the matter field change continuously from kink-like to double-kink-like solutions (Fig.\ref{fig3}). The transition happens specifically around $k\approx0.005$. This behavior is resounded in brane energy density with starts to show new energy peaks (Fig.\ref{fig4}). These peaks indicate the brane splitting. On the other hand, for $k<0$, changing the $k$-parameter leads to arising of other energy-critical points.

\begin{figure}[ht!]
\begin{center}
\begin{tabular}{ccc}
\includegraphics[height=4cm,width=5.2cm]{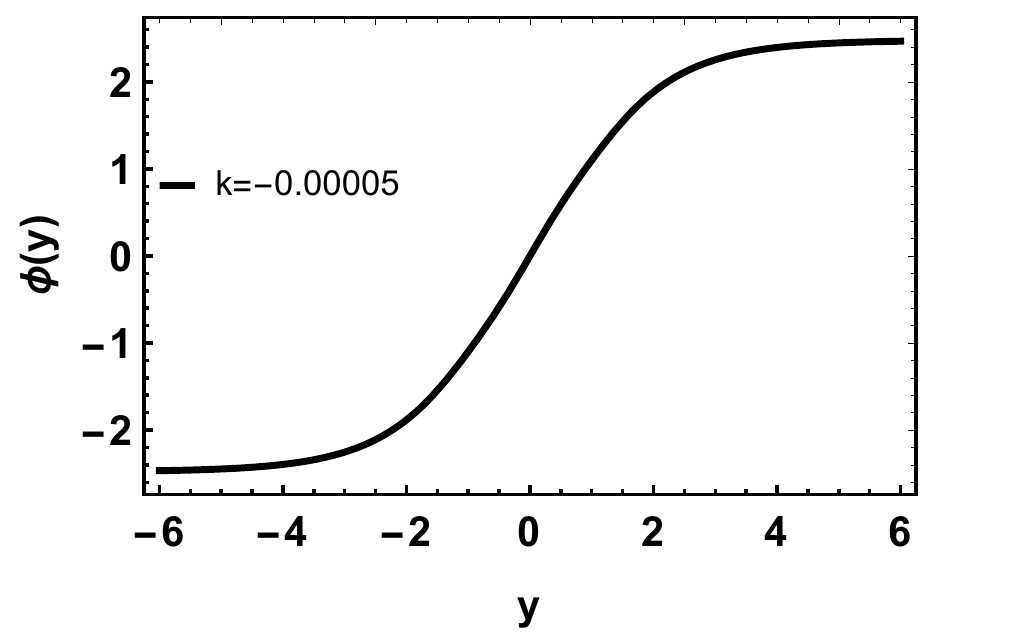}
\includegraphics[height=4cm,width=5.2cm]{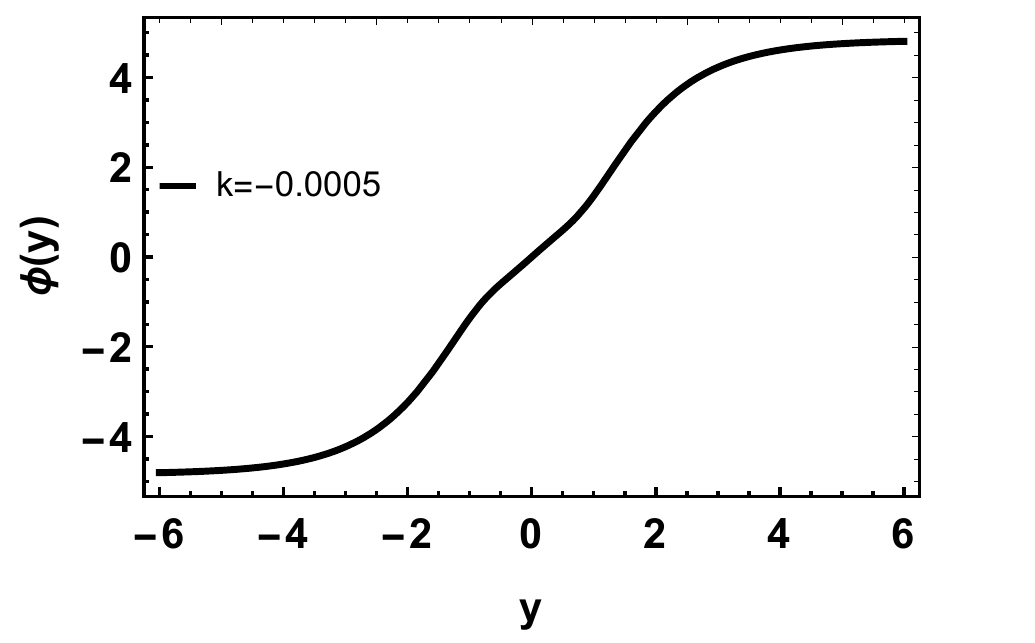}
\includegraphics[height=4cm,width=5.2cm]{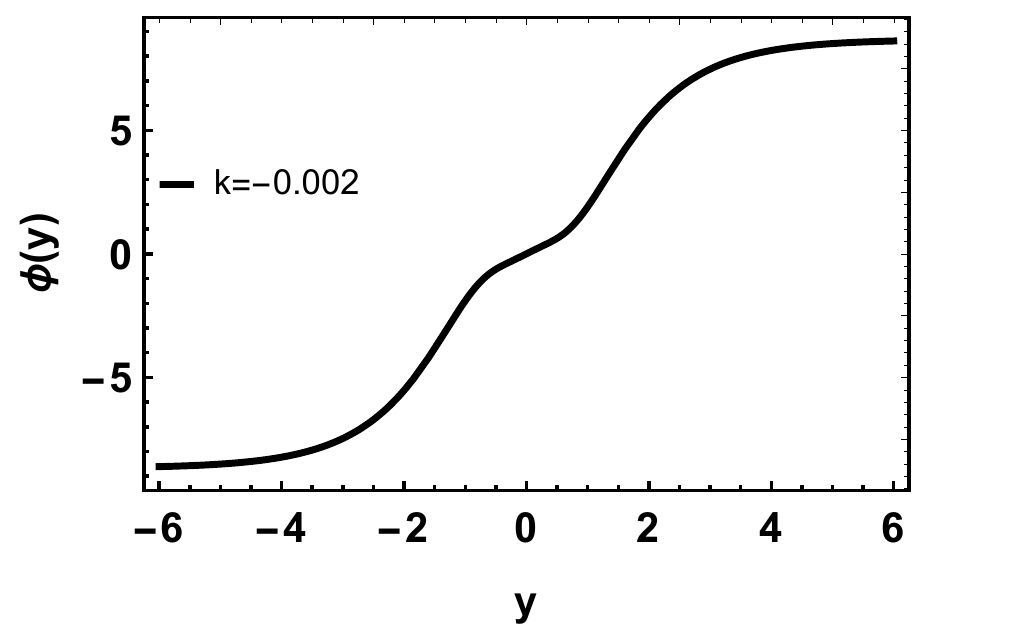}
\end{tabular}
\end{center}
\vspace{-0.5cm}
\caption{
Scalar field solution for $n=4$ with $p=\lambda=1$.
\label{fig5}}
\end{figure}

\begin{figure}[ht!]
\begin{center}
\begin{tabular}{ccc}
\includegraphics[height=5.5cm,width=7cm]{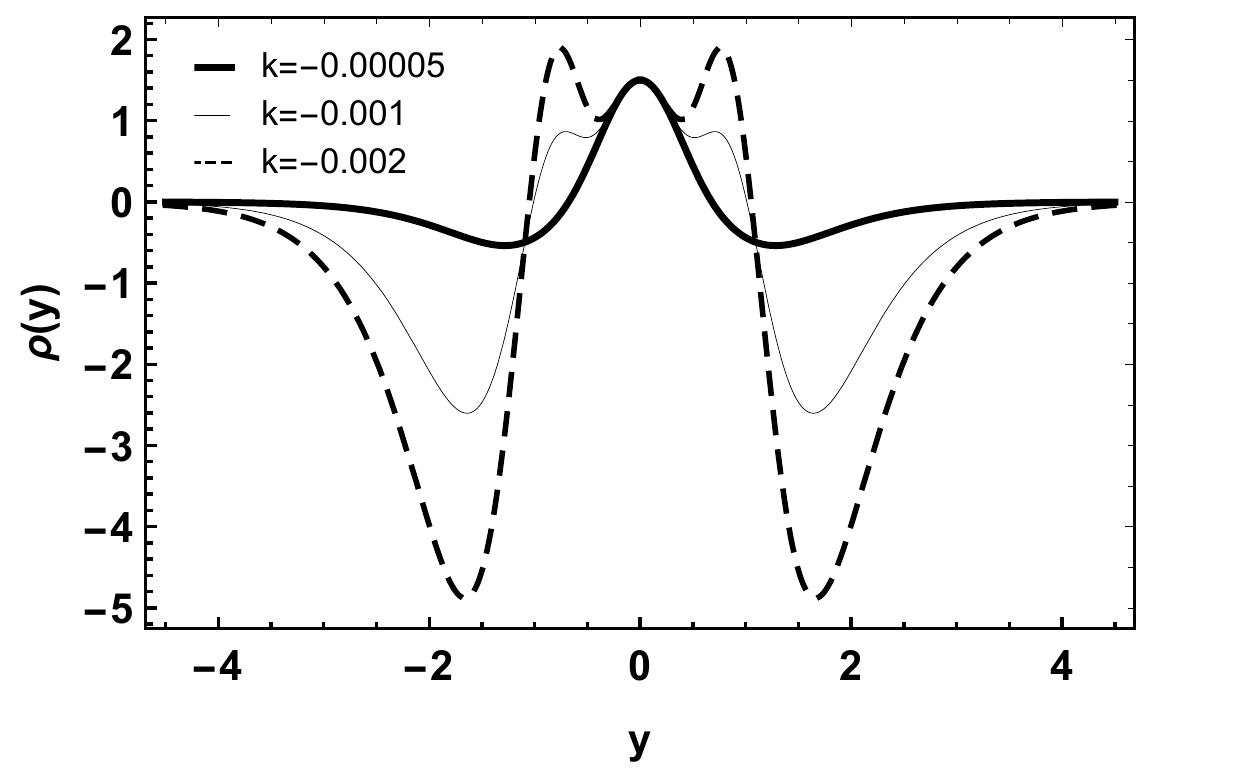}
\includegraphics[height=5.5cm,width=7cm]{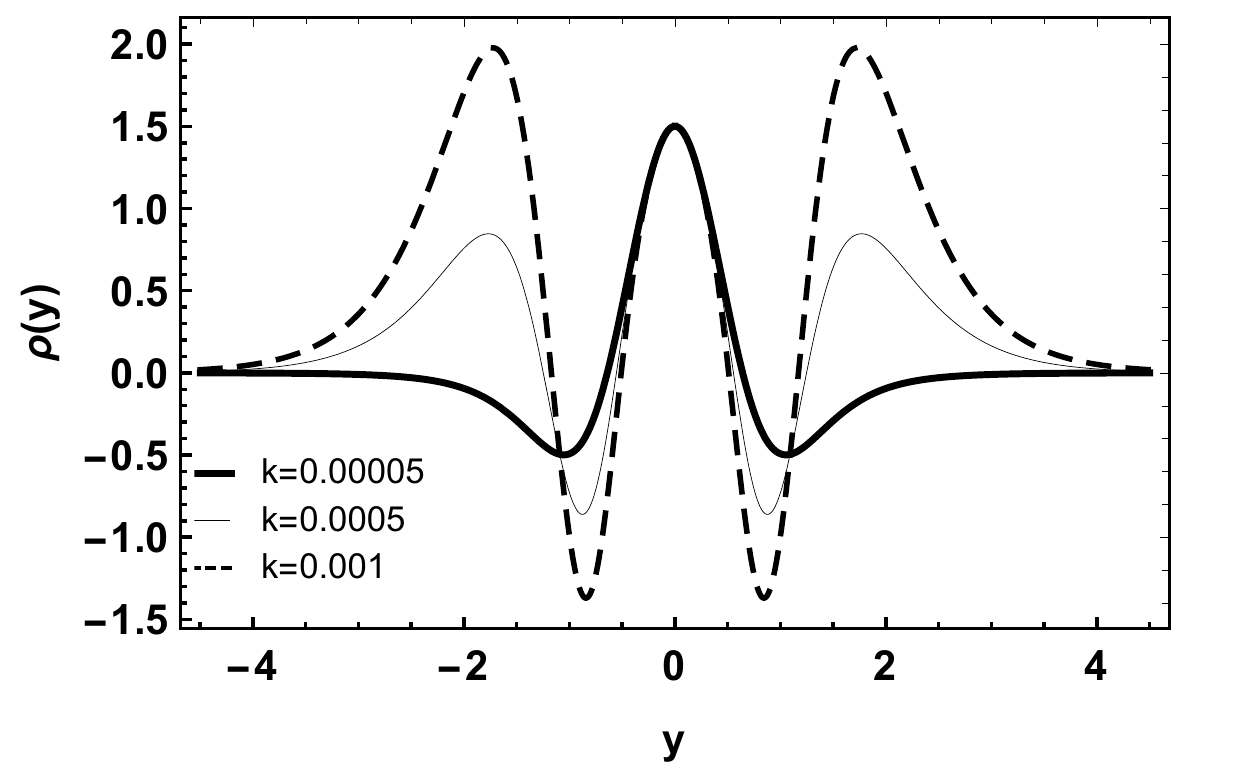}\\ 
(a) \hspace{7cm} (b)
\end{tabular}
\end{center}
\vspace{-0.5cm}
\caption{Brane energy density for $n=4$ with $p=\lambda=1$. (a) The case $k<0$ . (b) The case $k>0$.
\label{fig6}}
\end{figure}

Finally, for $n=4$ and $k<0$,  if $k$-parameter decrease the scalar field solution undergoes a alteration from a kink-like solution to a double-kink-like solution  (Fig. \ref{fig5}). This modification happens in the neighborhood of $k\approx-0.0004$. In this case, a brane splitting occurs with the arises of two energy-critical points (Fig. \ref{fig6}(a)). For $k>0$, the energy density presents the emergence of three peaks (see Fig. \ref{fig6}(b)).

For all values of $n$ and $k\to 0$, we restored the TEGR behavior. In this case, the scalar field has a kink-like solution. Thus, brane energy density will assume a profile well located with a single critical point near the localization of this solution \cite{DeWolfe1999, Gremm1999, Dzhunushaliev2009}.

For the $f_2(T)$ model, the equations (\ref{w.221}) and (\ref{w.222}), give us
\begin{align}\label{q.522}
\phi'^2(y)=\frac{3}{2}p\lambda^2\mathrm{sech}^2(\lambda y)\Big\{1-72p^2\lambda^2\tanh^{2}(\lambda y)\Big[\alpha-30\beta p^2\lambda^2\tanh(\lambda y)^{2}\Big]\Big\},
%\label{q.5522}
\end{align}

\begin{align}\nonumber
V(\phi(y))=&72p^6\lambda^7\mathrm{sech}^2(\lambda y)\tanh^{5}(\lambda y)\Big[\alpha-36\beta p^2\lambda^2\tanh^{2}(\lambda y)\Big]-432p^6\lambda^6\tanh^{6}(\lambda y)\times\\ \nonumber
&\Big[12(\alpha+\beta)p^2\lambda^2\tanh^{2}(\lambda y)-1\Big]^3-\frac{3}{2}p^2\lambda^3\tanh(\lambda y)\Big[\mathrm{sech}^2(\lambda y)+4p^2\lambda\tanh^{3}(\lambda y)\Big]\times\\
&\Big\{1-24p^2\lambda^4\tanh^{2}(\lambda y)\Big[\alpha-18\beta p^2\lambda^2\tanh^{2}(\lambda y)\Big]\Big\}.
\end{align}

Remembering the Eq. (\ref{3333}), we have the energy density is 
\begin{align}\nonumber
\rho(y)=&\frac{3\lambda^2}{2}\Bigg\{1-72p^2\lambda^2\tanh(\lambda y)^{2}\Big[\alpha-30\beta p^2\lambda^2\tanh(\lambda y)^{2}\Big]- 2p\sinh^2(\lambda y)\times\\
&\Bigg[1-36p^2\lambda^2\tanh(\lambda y)^{2}\Big[\alpha-20\beta p^2\lambda^2\tanh(\lambda y)^{2}\Big]\Bigg]\Bigg\}\cosh^{-2p}(\lambda y).
\end{align}

\begin{figure}[ht!]
\begin{center}
\begin{tabular}{ccc}
\includegraphics[height=4cm,width=5.2cm]{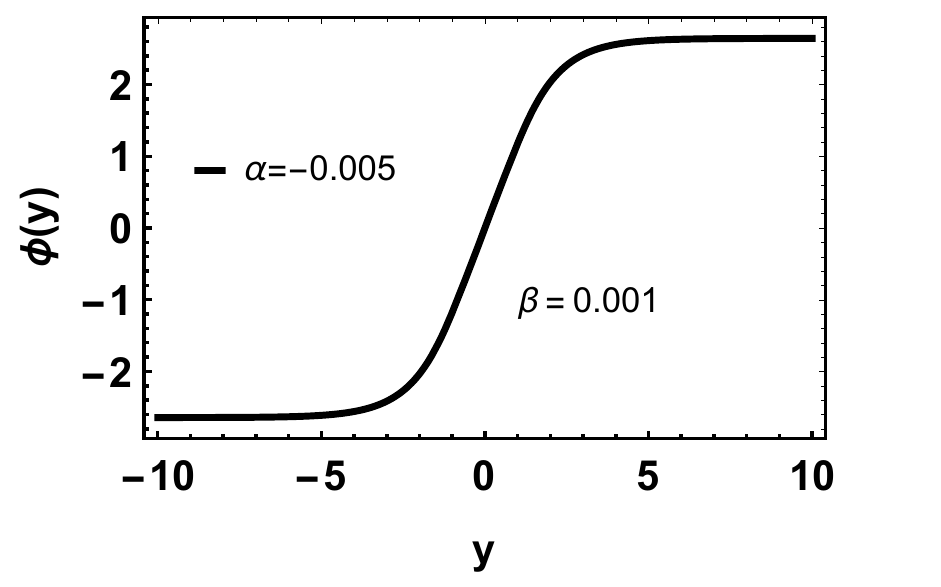}
\includegraphics[height=4cm,width=5.2cm]{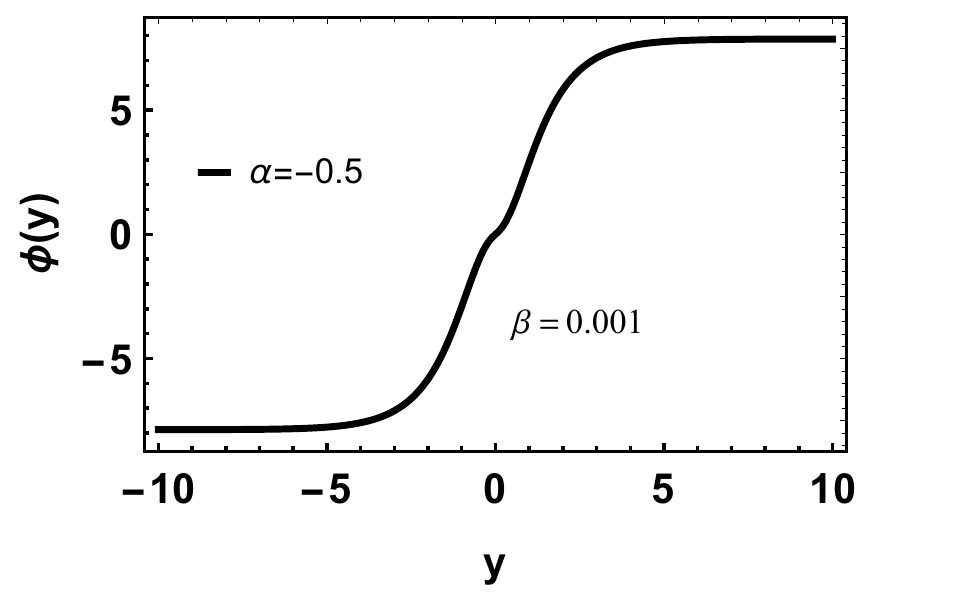}
\includegraphics[height=4cm,width=5.2cm]{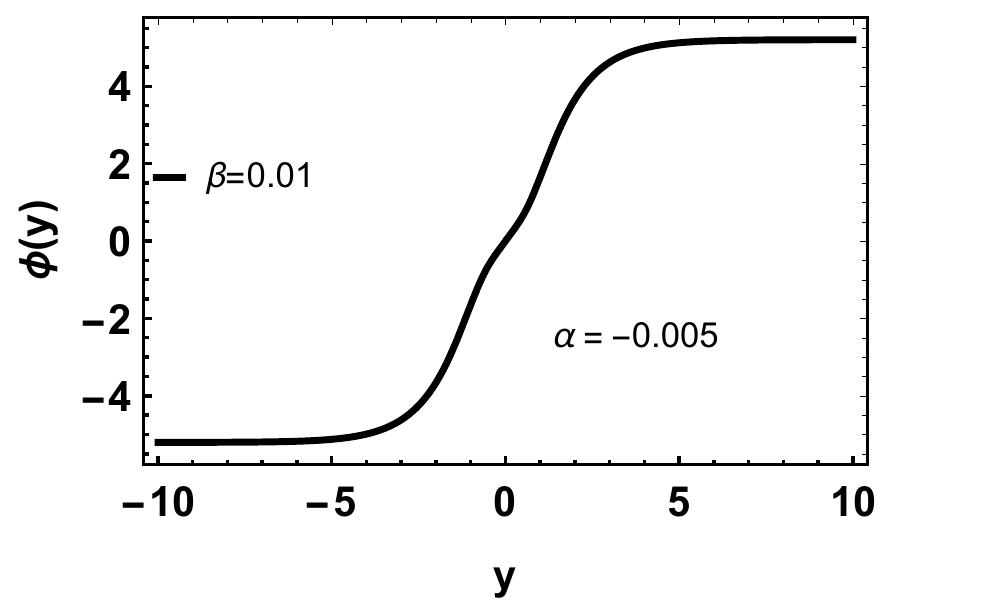}
\end{tabular}
\end{center}
\vspace{-0.5cm}
\caption{
Scalar field for the function $f_2(T)$ with $p=\lambda=1$.
\label{fig52}}
\end{figure}

\begin{figure}[ht!]
\begin{center}
\begin{tabular}{ccc}
\includegraphics[height=5.5cm,width=7cm]{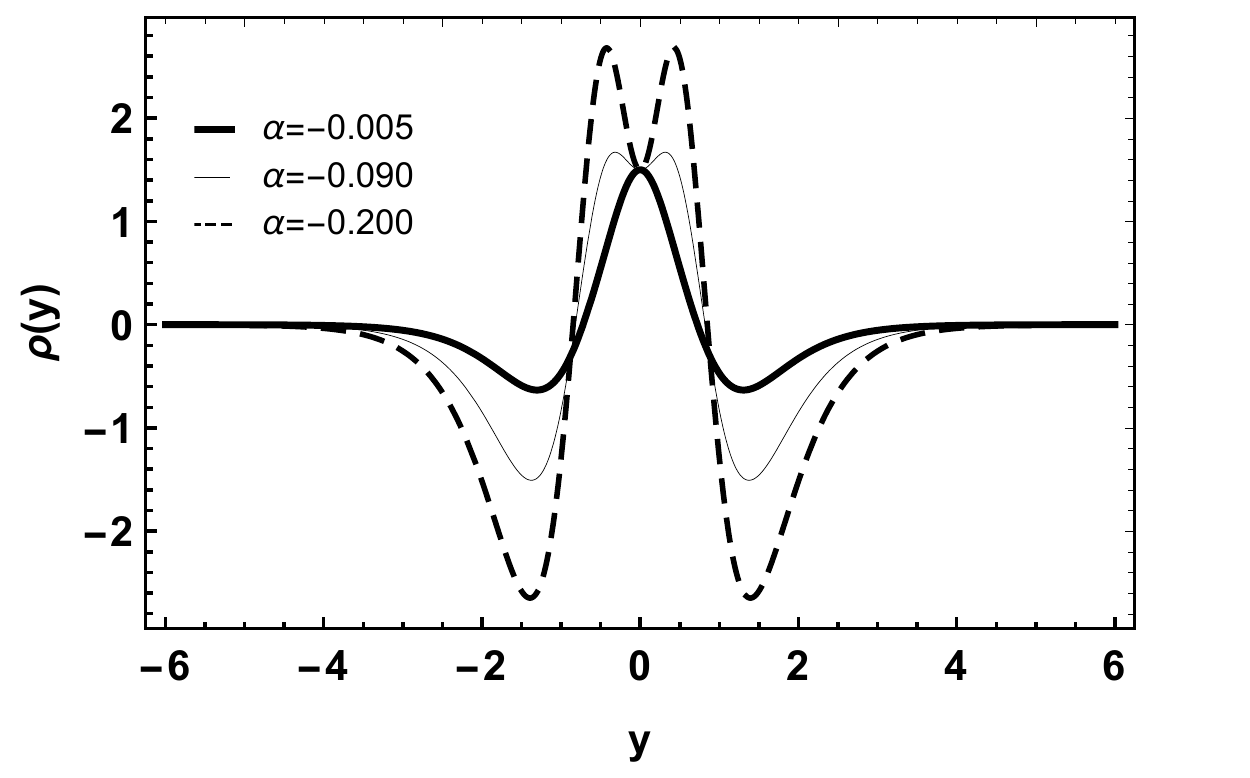}
\includegraphics[height=5.5cm,width=7cm]{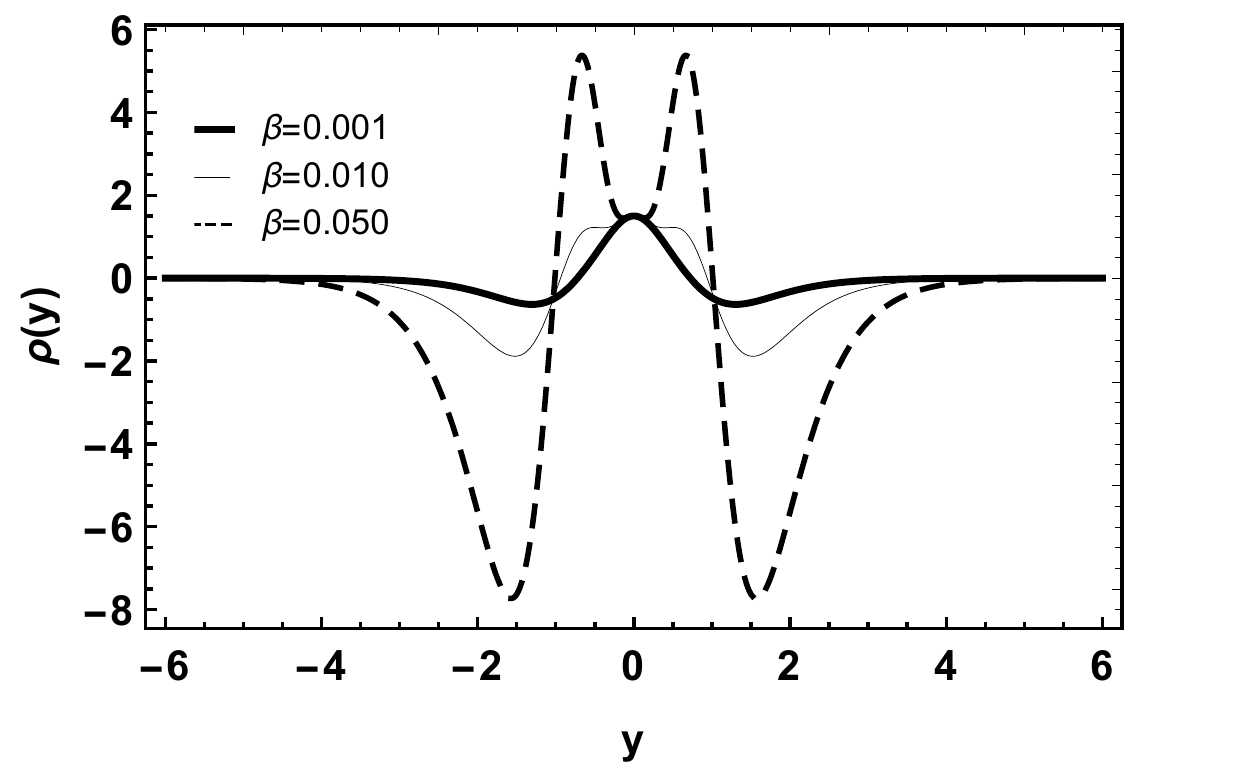}\\ 
(a) \hspace{7cm}(b)
\end{tabular}
\end{center}
\vspace{-0.5cm}
\caption{Brane energy density for the function $f_2(T)$ with $p=\lambda=1$. (a) The case $\beta=0.001$. (b) The case $\alpha=-0.005$.
\label{fig62}}
\end{figure}

In the case $f_2(T)$, we show the numerical solution of the matter field in Fig. \ref{fig52}. In this case, when the $\alpha$-parameter decreases, the scalar field changes from a kink-like to a double-kink configuration. The behavior of the $f_2(T)$ model seems to be similar to the $f_1(T)$ model but with a smoother field deformation. An analogous deformation occurs when the $\beta$-parameter increases. By numerical calculation was perceived that when $\alpha\approx-0.05$ and $\beta\approx0.005$, the deformation of the fields starts to happen. At that moment, in the energy density (Fig.\ref{fig62}), two critical points (peaks) appear, indicating the brane splitting. Besides, the GR is retrieved again when $\alpha, \beta\to 0$ \cite{DeWolfe1999,Gremm1999,Dzhunushaliev2009}. In this case, the matter field will assume a kink-like profile.

\subsection{The case $A_2(y)=\ln\vert\tanh[\lambda(y+c)]-\tanh[\lambda(y-c)]\vert$}

Analyzing the case $A_2(y)=\ln\vert\tanh[\lambda(y+c)]-\tanh[\lambda(yc)]\vert$, we notice that the alteration of the parameters $k$ and $n $ lead us to a similar behavior of the matter field and the energy density of the model $A_1(y)=-p\ln{\cosh(\lambda y)}$. Therefore, for the $A_2(y)$ model, allow us to turn our attention to the study of the changes generated by changing of the $\lambda$ and $c$ parameters.

\begin{figure}[ht!]
\begin{center}
\begin{tabular}{ccc}
\includegraphics[height=5.5cm,width=7cm]{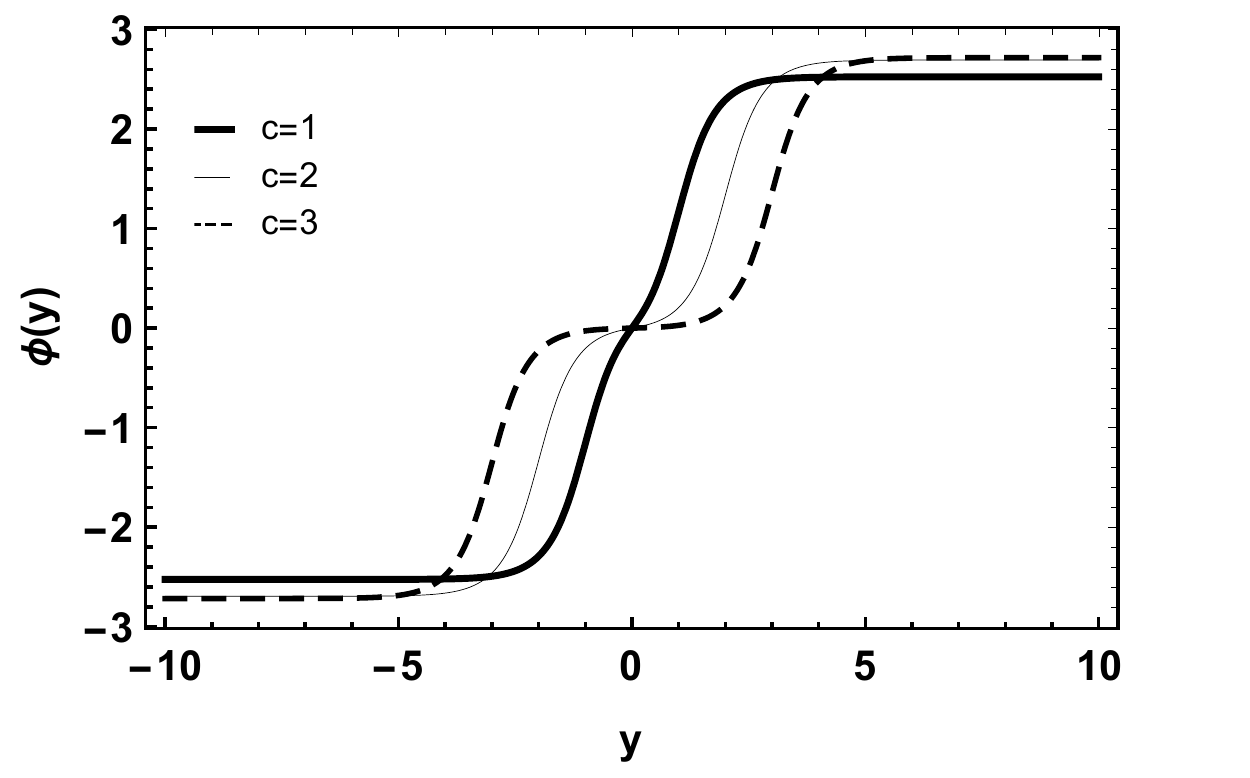}
\includegraphics[height=5.5cm,width=7cm]{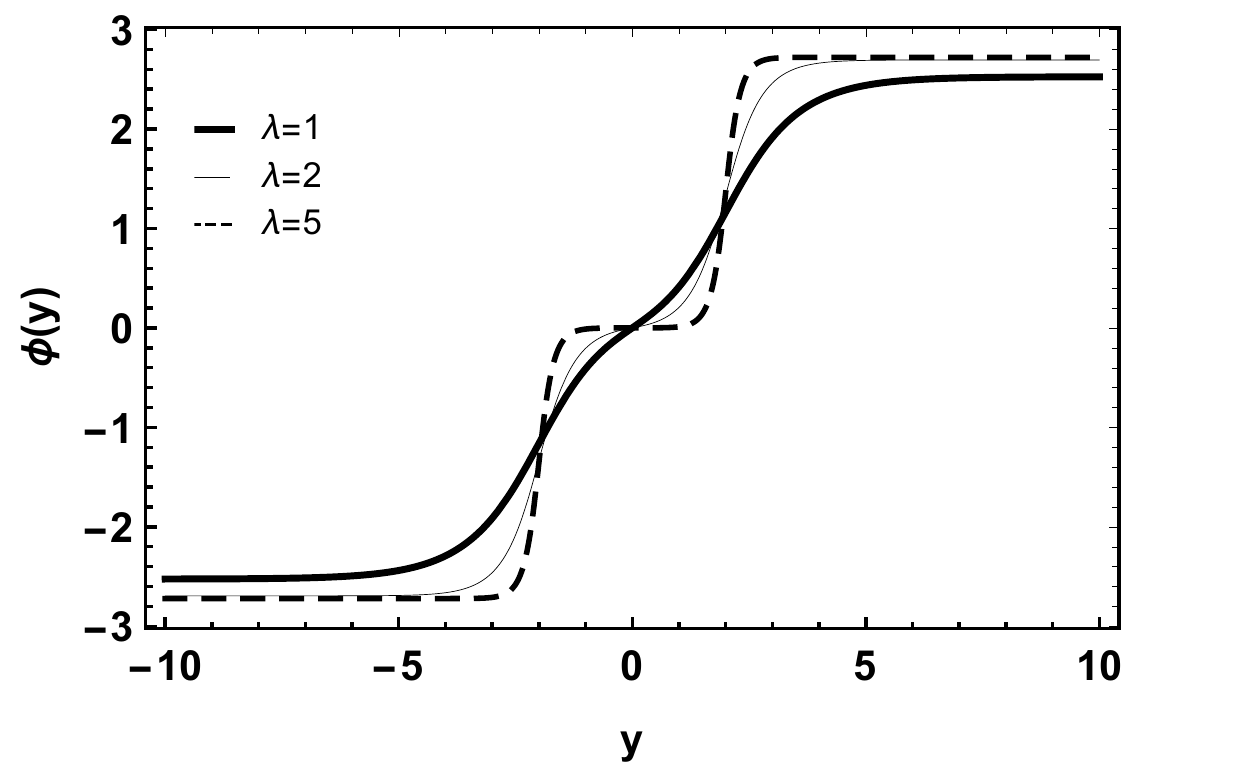}\\ 
(a) \hspace{7cm} (b)
\end{tabular}
\end{center}
\vspace{-0.5cm}
\caption{Scalar field solution for the $f_1(T)$ function with $n=1$ and $k=-0.5$. (a) The case $\lambda=2$ . (b) The case $c=2$.
\label{fig63}}
\end{figure}

An interesting result happens when the case $A_2(y)$ is analyzed. In this case, the scalar field solution shown in Fig. \ref{fig63} has a compact-like profile. Matter fields with profile compact-like were studied in several scenarios, such as in low-dimensional topological theories and braneworld, see e. g., Ref. \cite{LA1}. The compact-like solutions are structures described by a field that reaches the vacuum expected value (VEV) in a finite region and has the shape shown in Fig. \ref{fig63}. It is worth mentioning that the compact-like configurations only appear in our model when the $\lambda$-parameter increases. Indeed, when $\lambda=5$ and $c=2$, we have configurations that appear to be double-compact configurations.

\begin{figure}[ht!]
\begin{center}
\begin{tabular}{ccc}
\includegraphics[height=5.5cm,width=7cm]{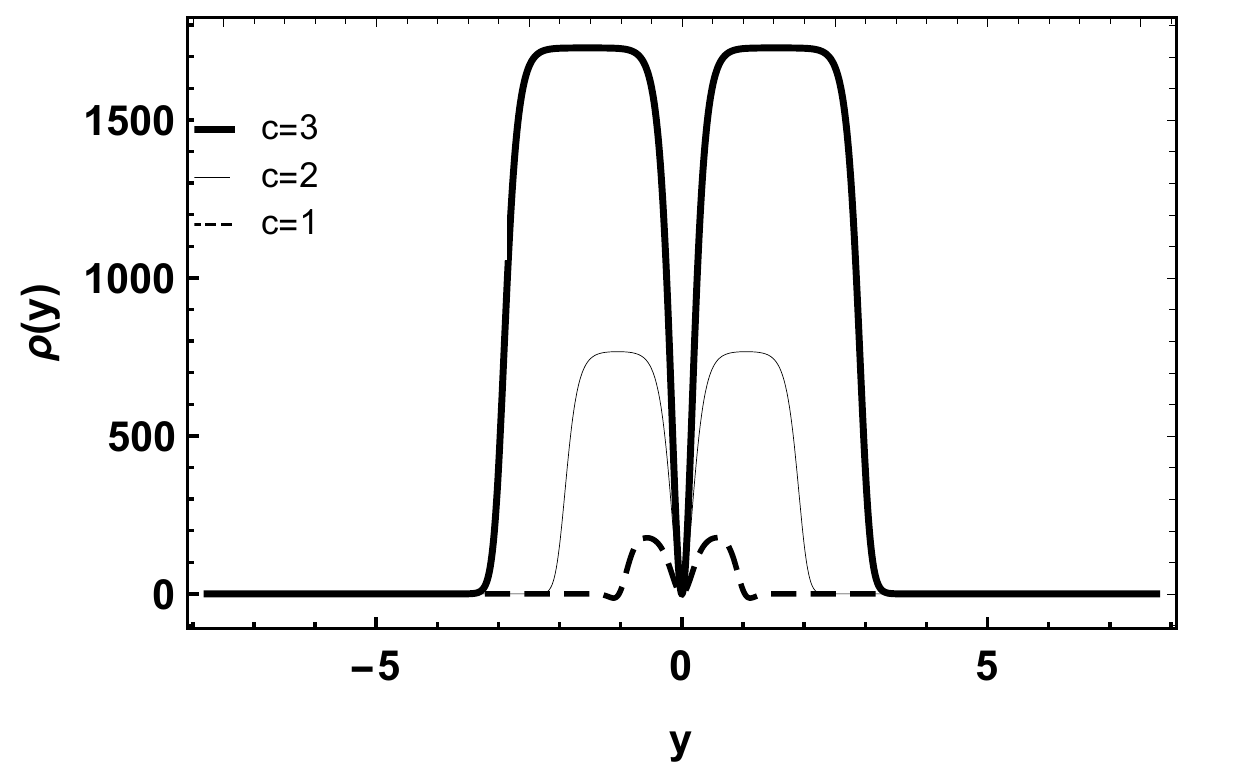}
\includegraphics[height=5.5cm,width=7cm]{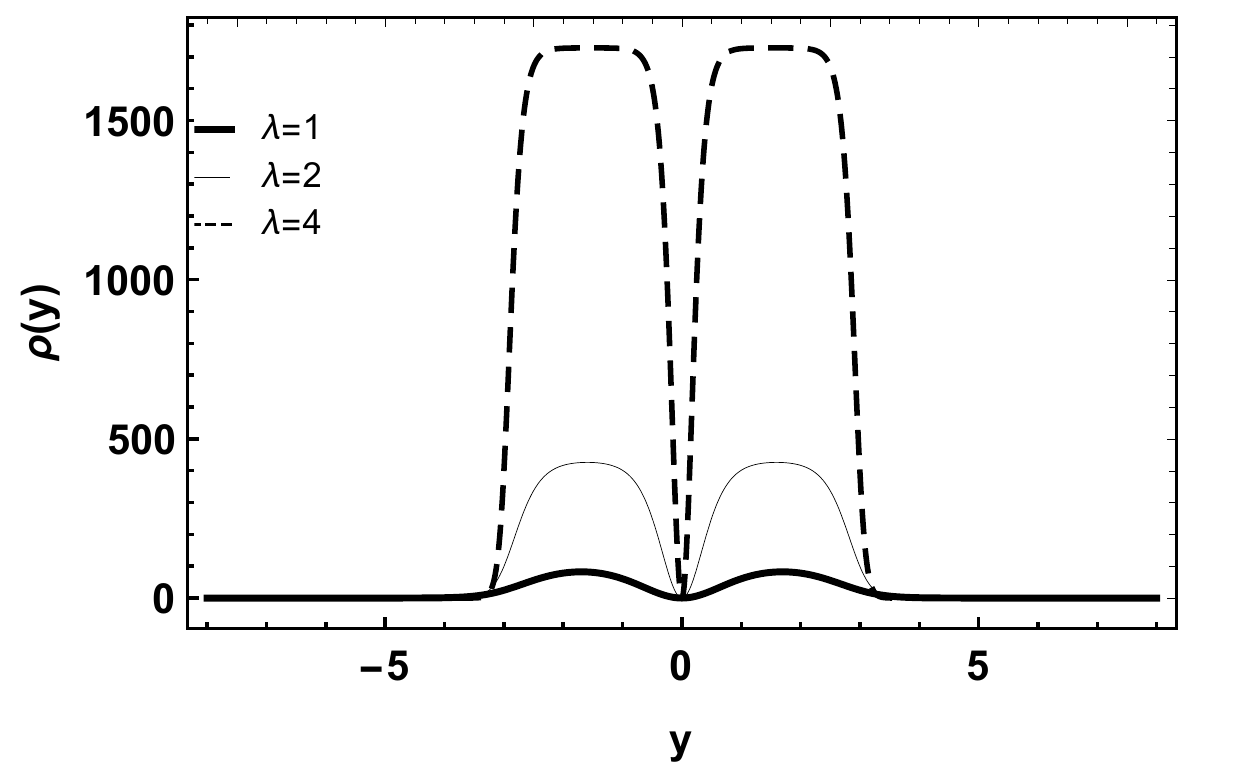}\\ 
(a) \hspace{7cm} (b)
\end{tabular}
\end{center}
\vspace{-0.5cm}
\caption{Brane energy density for $f_1(T)$ with $n=1$ and $k=-0.5$. (a) The case $\lambda=2$. (b) The case $c=2$.
\label{fig622}}
\end{figure}

In Fig. \ref{fig622}, is shown the behavior of the brane energy density. Note that the energy density is the mirror of the profile of the scalar field. Indeed, when are obtained the compact-like profiles, the energy density happens to have the shape of a box or a profile type the Heaviside-like function. This energy profile confirms our preliminary hypothesis that the matter field solutions are compact-like solutions in this model.

%%%%%%%%%%%%%%%%%%%%%%%%%%%%%%%%%%%%%%%%%%%%%%%%%%%%%%%%%%%%%%%%%%%%%%%%%%%%%%%%%
\section{DCE: Phase transition and brane stability}
\label{sec2}

Let us use the Differential Configurational Entropy (DCE) to analyze in detail our results. In Ref. \cite{GS}, Gleiser and Stamatopoulos (GS) define the configurational entropy (CE) as a representation of detailed measurements of the complexity of fields. DCE, a variant of the CE, has shown significant results in analyzing the informational content of localized structures in several models, see Refs. \cite{Chinaglia2017,Correa2015c, Correa2016b}. Therefore, we are encouraged to build this section a brief review of the DCE concept. This discussion is important for our analysis of the phase transitions in braneworld with $f(T)$ gravity.

It is defined the DCE by the Fourier transform of the energy density, i. e.,
\begin{align}\label{4444}
\mathcal{F}[\omega]=\frac{1}{\sqrt{2\pi}}\int e^{i\omega y}\rho(y) dy.
\end{align}
Here, the energy density is $\rho(y)=-e^{2A}\mathcal{L}_m$ (see Eq. (\ref{3333})). Adopting the definition of the Eq. (\ref{4444}), the Fourier transform is rewritten as
\begin{align}
\mathcal{F}[\omega]=-\frac{1}{\sqrt{2\pi}}\int e^{2A(y)+i\omega y}\mathcal{L}_m dy.
\end{align}

To define the DCE, let us first build the modal fraction as
\begin{align}
f(\omega)=\frac{\mid\mathcal{F}[\omega]\mid^2}{\int\mid\mathcal{F}[\omega]\mid^2d\omega}.
\end{align}
The modal fraction is defined $\leq 1$. Indeed, the modal fraction is the relative weight of each $\omega$ mode.

For the case of localized and continuous function $f(\omega)$, the DCE is defined as \cite{Chinaglia2017,GS,Correa2015a,G2,Correa2015b,Correa2015c,Correa2016b}
\begin{align}\label{dce_g}
S_C[f]=-\int \bar{f}(\omega)\ln[\bar{f}(\omega)]d\omega,
\end{align}
where $\bar{f}(\omega)=f(\omega)/f_{max}(\omega)$ is the normalized modal fraction, and $f_{max}(\omega)$ is the maximum value of the fraction.

Since the DCE is defined, we can apply this tool to analyze our $f(T)$ braneworld scenario.

\subsection{The case $A_1(y)=-p\ln{\cosh(\lambda y)}$}

For the $f_1(T)$ model when $n=2$, the modal fraction is
\begin{align}
f(\omega)=\frac{231\pi\omega^2[5\omega^2+2k(48-80\omega^2+7\omega^4)]^2}{640[55+96k(148k-11)]}\mathrm{csch}^2\Big(\frac{\pi\omega}{2}\Big).
\end{align}

\begin{figure}[ht!]
\begin{center}
\begin{tabular}{ccc}
\includegraphics[height=5.5cm,width=7cm]{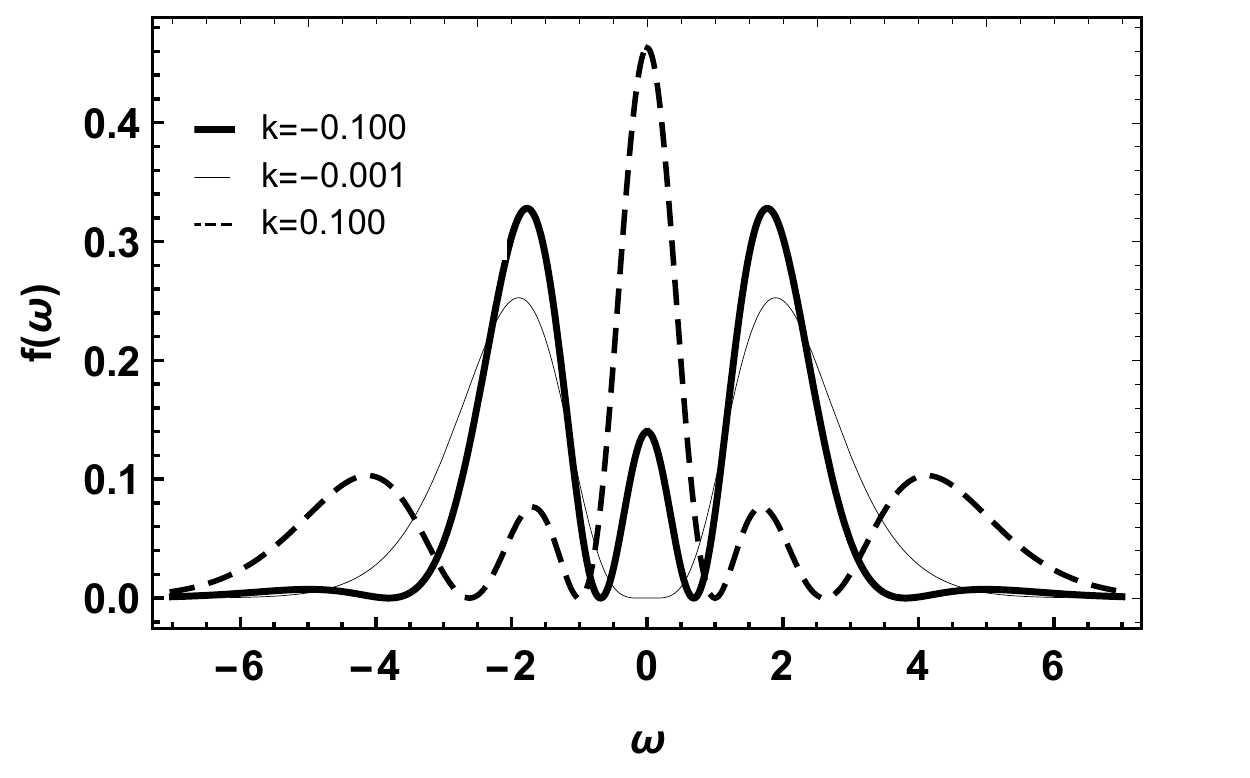}
\includegraphics[height=5.5cm,width=7cm]{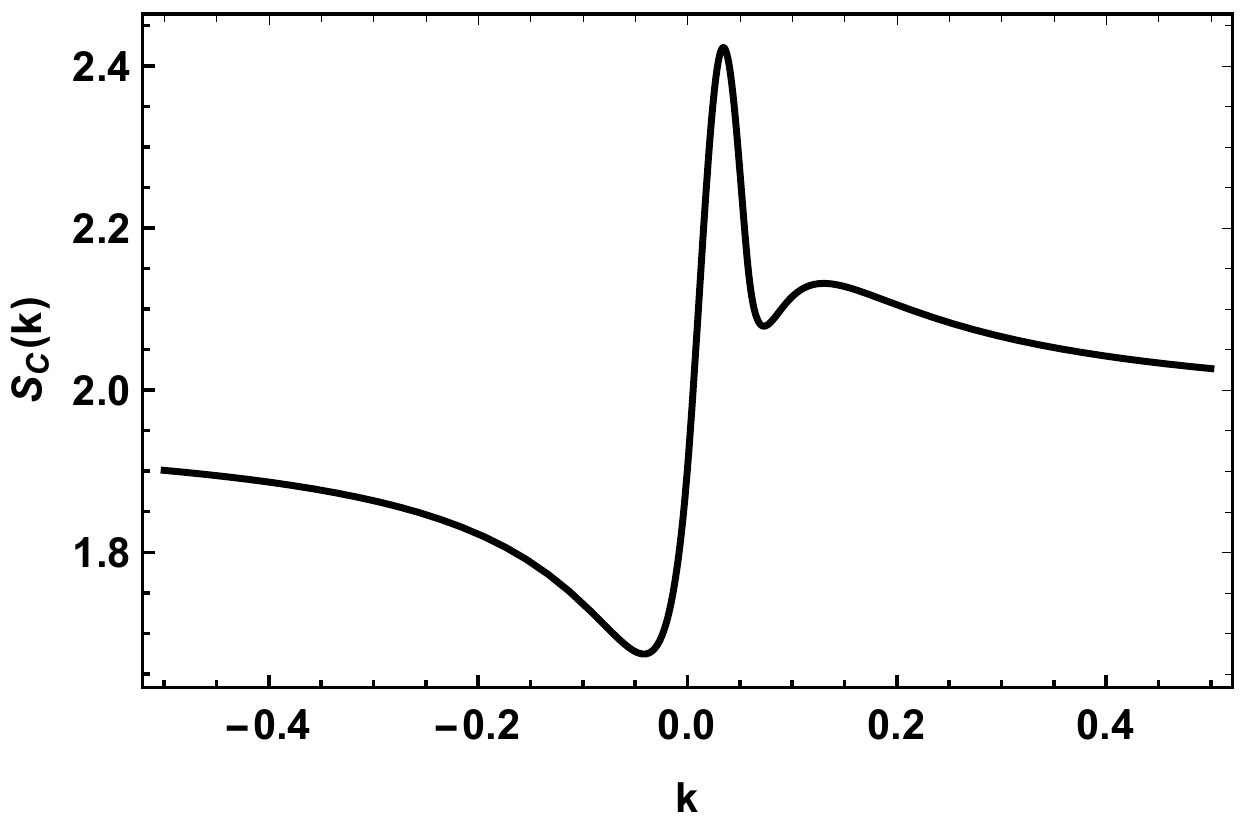}\\ 
(a) \hspace{7cm} (b)
\end{tabular}
\end{center}
\vspace{-0.5cm}
\caption{(a) Modal fraction with $n=2$ and $p=\lambda=1$. (b) DCE with $n=2$ and $p=\lambda=1$.
\label{fig7}}
\end{figure}

In Fig. \ref{fig7}, the modal fraction (Fig. \ref{fig7}(a)) and the DCE (Fig. \ref{fig7}(b)) of the model are shown for $n =2$. The DCE can only be analyzed numerically. In this case, DCE has two minimum points and one maximum point. The absolute minimum that appears in DCE is located in the range $-0.05<k<-0.03$. In this range of values, there is a transition from kink-like solutions to double-kink solutions. We interpreted this transition as an evolution of a single phase to a double phase transition.

For $n=3$, the modal fraction is
\begin{align}
f(\omega)=\frac{429\pi\omega^2[35\omega^2+18k(952\omega^2-480-140\omega^4+3\omega^6)]^2}{4480[715+1728k(15624k+65)]}\mathrm{csch}^2\Big(\frac{\pi\omega}{2}\Big).
\end{align}

\begin{figure}[ht!]
\begin{center}
\begin{tabular}{ccc}
\includegraphics[height=5.5cm,width=7cm]{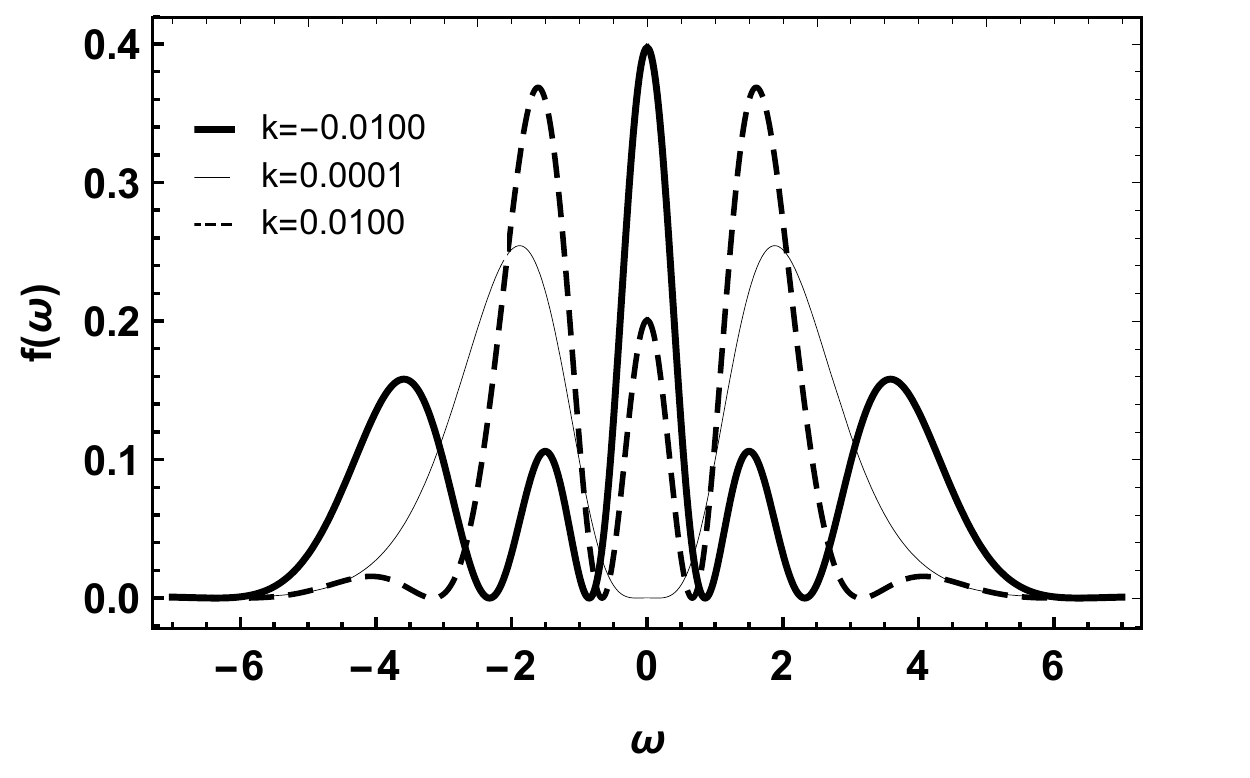}
\includegraphics[height=5.5cm,width=7cm]{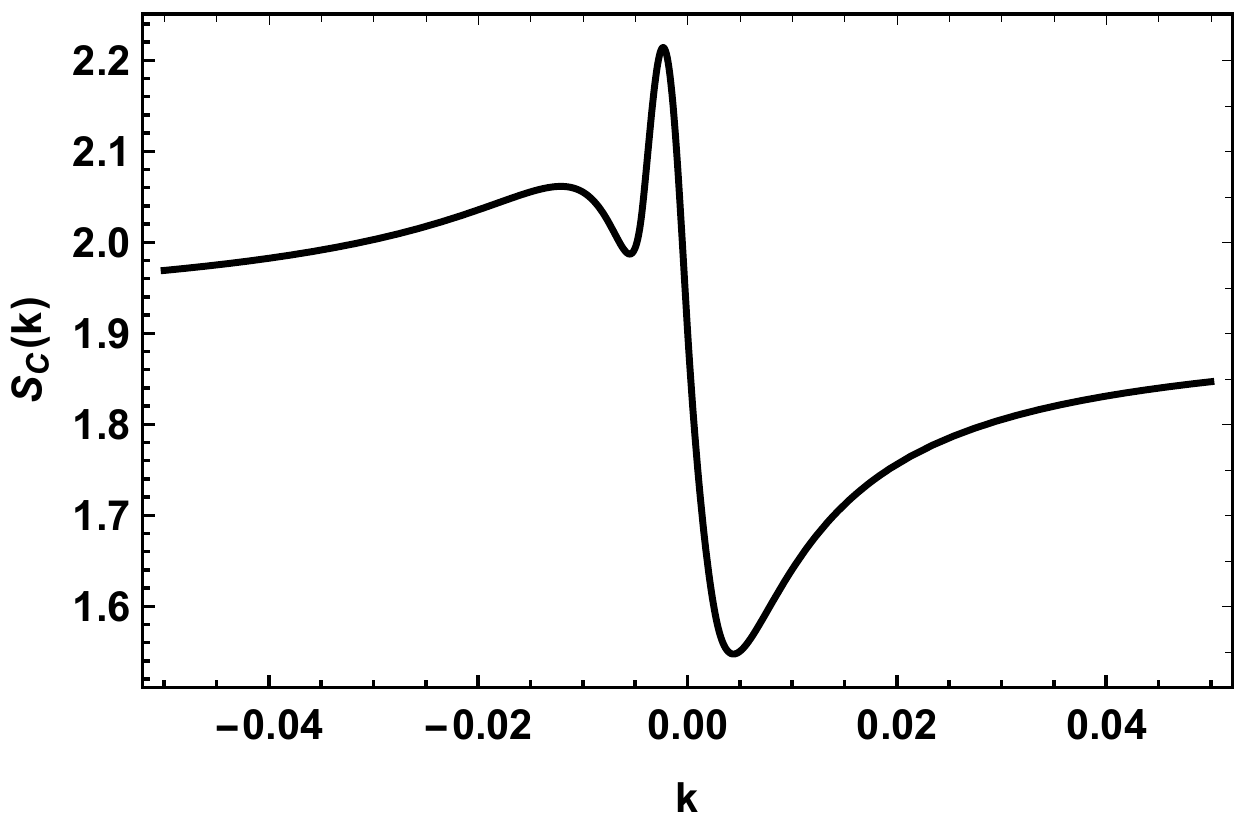}\\ 
(a) \hspace{7cm} (b)
\end{tabular}
\end{center}
\vspace{-0.5cm}
\caption{(a) Modal fraction with $n=3$ and $p=\lambda=1$. (b) DCE with $n=3$ and $p=\lambda=1$.
\label{fig8}}
\end{figure}

In Fig. \ref{fig8}, the modal fraction (Fig. \ref{fig8}(a)) and the DCE (Fig. \ref{fig8}(b)) are shown for $n=3$. Analogously to the case $n=2$, the DCE has two minimum points and one maximum point. The absolute minimum point with values at the range $0.003<k<0.005$ again suggests the appearance of a double phase transition. This double phase transition accompanies the appearance of internal structures and the brane splitting.

Finally, for $n=4$ the modal fraction is
\begin{align}
f(\omega)=&\frac{46189\pi\omega^2[105\omega^2+4k(80640-180736\omega^2+34944\omega^4-1344\omega^6+11\omega^8)]^2}{13440[230945+193536k(6576768k-1615)]}\text{csch}^2\Big(\frac{\pi\omega}{2}\Big).
\end{align}

\begin{figure}[ht!]
\begin{center}
\begin{tabular}{ccc}
\includegraphics[height=5.5cm,width=7cm]{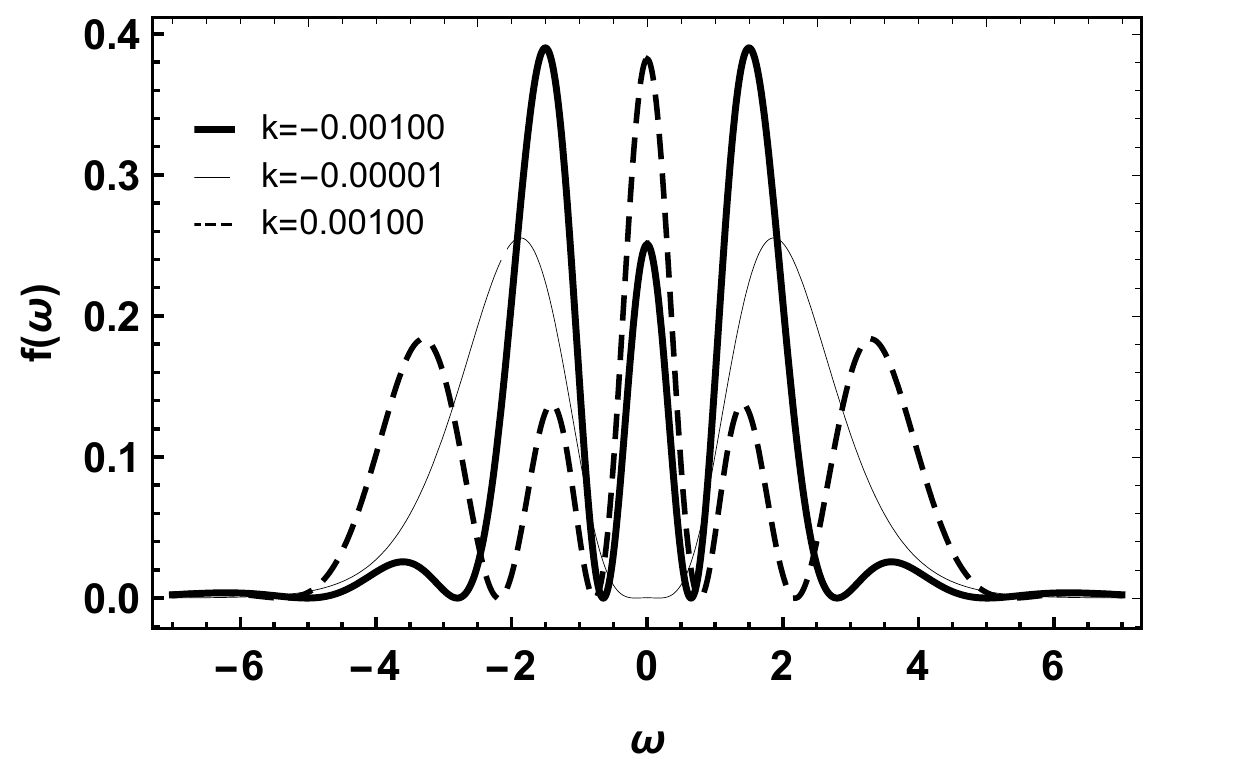}
\includegraphics[height=5.5cm,width=7cm]{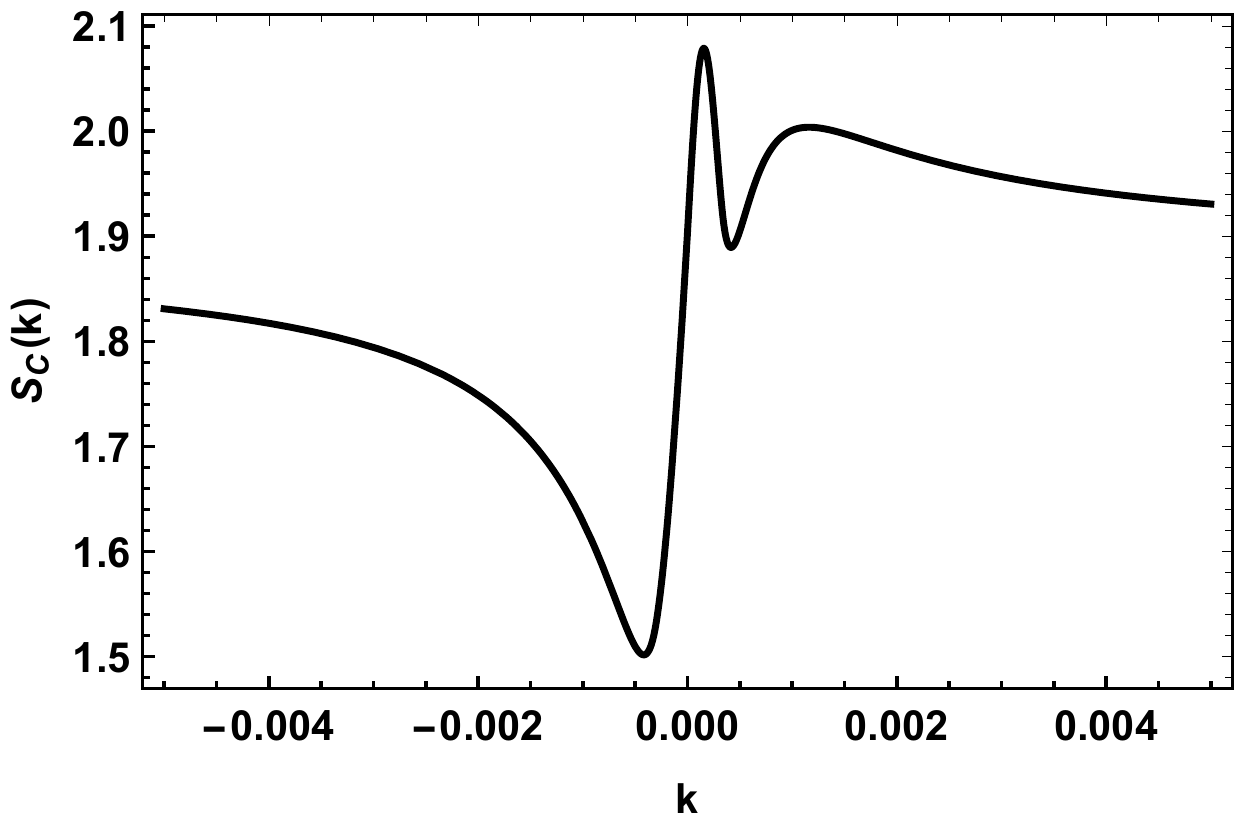}\\ 
(a) \hspace{7cm} (b)
\end{tabular}
\end{center}
\vspace{-0.5cm}
\caption{(a) Modal fraction with $n=4$ and $p=\lambda=1$. (b) DCE with $n=4$ and $p=\lambda=1$.
\label{fig9}}
\end{figure}

In Fig. \ref{fig9}, the modal fraction [Fig. \ref{fig9}(a)] and the DCE [Fig. \ref{fig9}(b)] are exhibited for $n=4$. Again, the DCE has two minimum points and one maximum point. In this case, the absolute minimum point is located in the range $-0.0005<k<-0.0003$ indicating the multiple phase transition, the emergence of internal structures, and brane splitting.

Let us now connect the discussion from the Sec. \ref{sec1} to the results obtained in Sec. \ref{sec2} for the modal fraction. Perceive that when we vary k, the modal fraction tends to have several oscillations. However, for $n=2$ and $k=-0.001$, the modal fraction presents two peaks of greater intensity, establishing a range of values that the field reaches the stability. On the other hand, when $k=-0.1$, the new peaks indicate the range of values where the kink-like transforms into a double-kink-like configuration. This result from the DCE announces the measure of the $\omega$  parameters (in power spectrum) that the phase modification and brane splitting occur.

Similar behavior happens for $n=3$, i. e. when $k=0.0001$, the modal fraction presents two peaks of greater intensity near the range of values of $k$ that the scalar field reaches its stability. For $k=0.01$, the new peaks indicate the interval at the power spectrum where the kink-like transforms into a double-kink-like configuration. The same goes for $n=4$.

Analyzing the DCE results, we perceived that the results indicate the stability points in the solutions of the matter field. In other words, the DCE tells us the value of model parameters that are most likely to find a stable field configuration. For n=2, the stability point appears at $k\approx -0.05$. Meanwhile, for $n=3$, the stability point is at $k\approx 0.005$. Finally, for $n=4$, the stability point is at $k\approx -0.0004$.

For the case $f_2(T)$, the modal fraction is
\begin{align}
f(\omega)=&\frac{429\pi\omega^2[35\omega^2+84\alpha(4-10\omega^2+\omega^4)+10\beta(1400\omega^2-224\omega^4+5\omega^6-576)]^2}{4480[715+106704\alpha^2+81120\beta+1814400\beta^2-72\alpha(143+34800\beta)]}\text{csch}^2\Big(\frac{\pi\omega}{2}\Big).
\end{align}

\begin{figure}[ht!]
\begin{center}
\begin{tabular}{ccc}
\includegraphics[height=5.5cm,width=7cm]{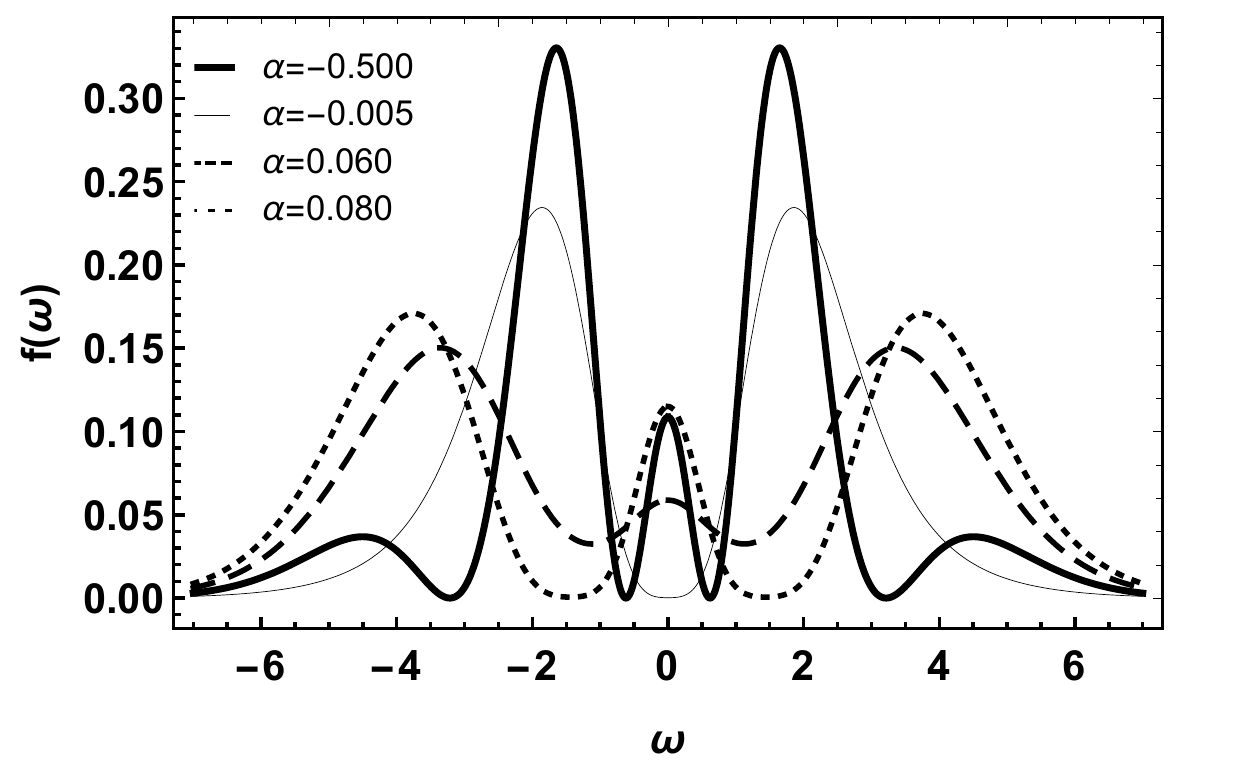}
\includegraphics[height=5.5cm,width=7cm]{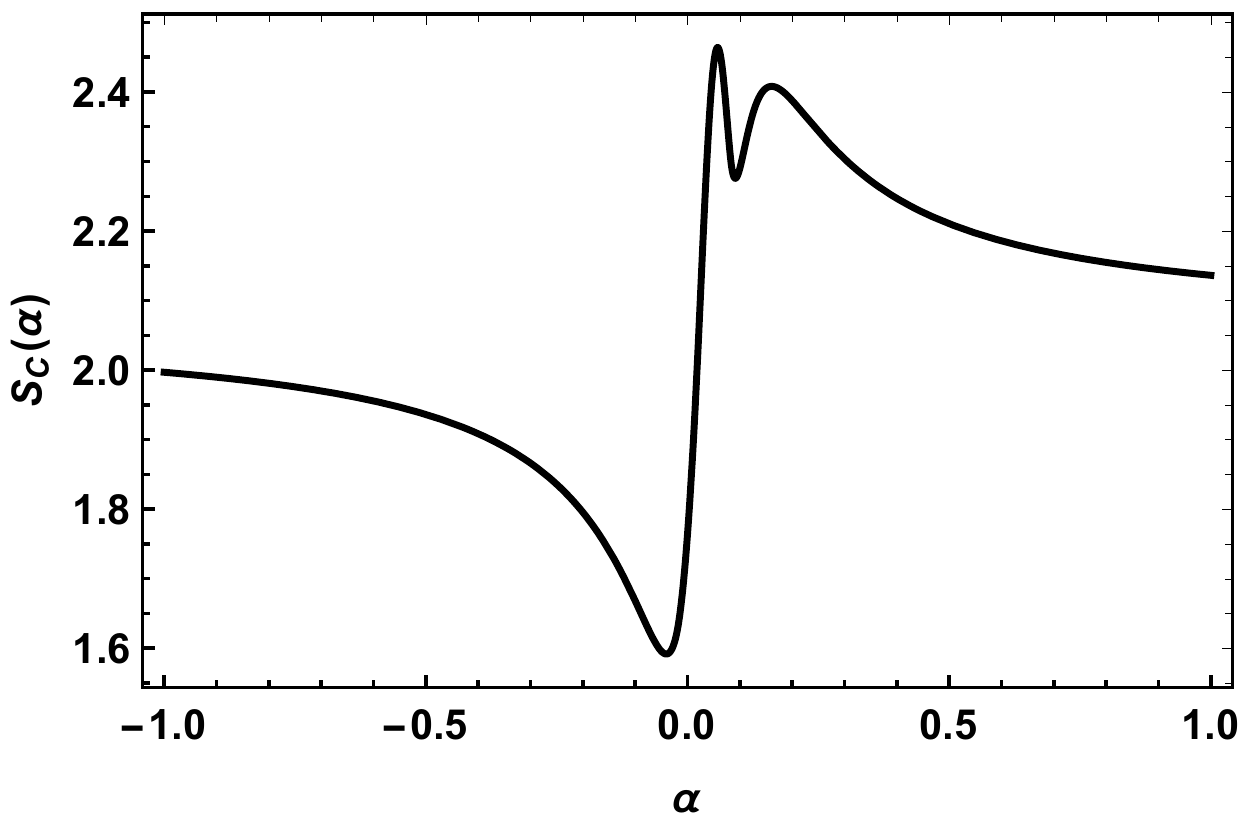}\\ 
(a) \hspace{7cm}(b)\\
\includegraphics[height=5.5cm,width=7cm]{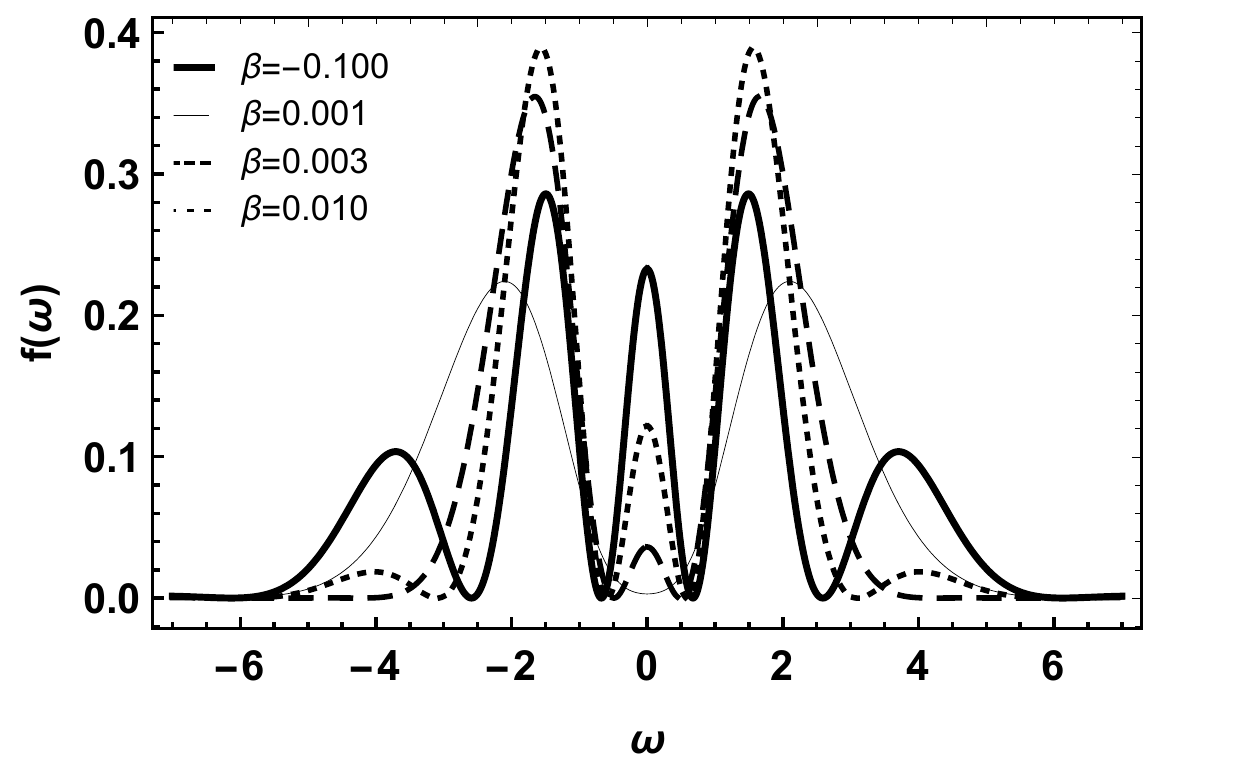}
\includegraphics[height=5.5cm,width=7cm]{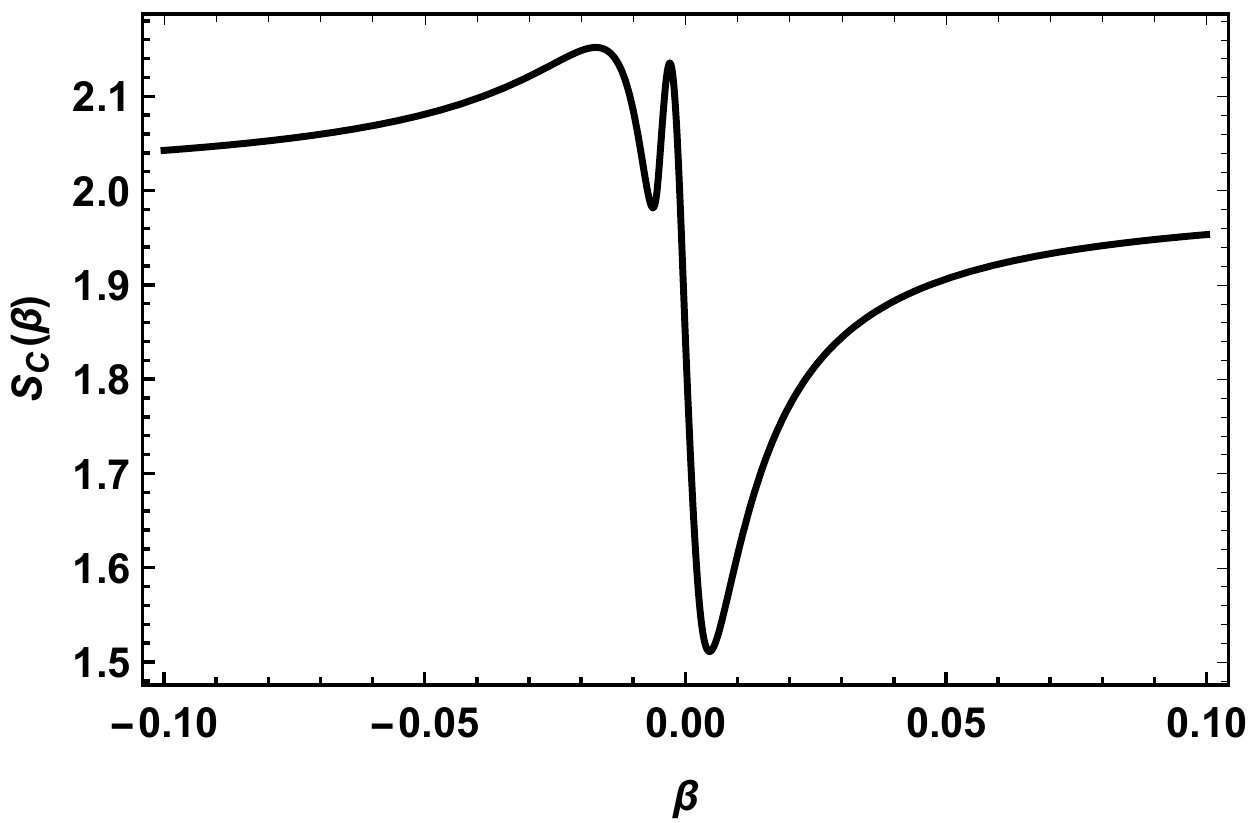}\\ 
(c) \hspace{7cm} (d)
\end{tabular}
\end{center}
\vspace{-0.5cm}
\caption{(a) Modal fraction of the $f_2(T)$ model with $p=\lambda=1$ and $\beta=0.001$. (b) DCE with $p=\lambda=1$ and $\beta=0.001$.
(c) Modal fraction of the $f_2(T)$ model with $p=\lambda=1$ and $\alpha=-0.005$. (c) DCE with $p=\lambda=1$ and $\alpha=-0.005$.}
\label{fig92}
\end{figure}

In Fig. \ref{fig92}, the modal fraction [Fig. \ref{fig92}(a) and \ref{fig92}(b)] and the DCE [Fig.\ref{fig92} (c)] are traced for the case $f_2(T)$. Note that the DCE has an absolute minimum point located in the range $-0.2<\alpha<-0.005$ and $0.001<\beta<0.05$. 
 
The modal fraction profile changes as we vary the parameter $\alpha$ [see Figs. \ref{fig92}(a) and \ref{fig92}(c)]. Note that if $\alpha=-0.05$ and $\beta=0.001$ the modal fraction presents a configuration of smaller oscillation. In this configuration, there are two peaks near the values of the parameters $\alpha$ and $\beta$ that are used for finding the scalar field obtained in Sec. \ref{sec1}. The results from the DCE allow us to affirm that stable field configurations are more likely to be found around $\alpha\approx-0.05$ and $\beta\approx0.005$ [see Figs. \ref{fig92}(b) and \ref{fig92}(d)].

\subsection{The case $A_2(y)=\ln\vert\tanh[\lambda(y+c)]-\tanh[\lambda(y-c)]\vert$}

\begin{figure}[ht!]
\begin{center}
\begin{tabular}{ccc}
\includegraphics[height=5.5cm,width=7cm]{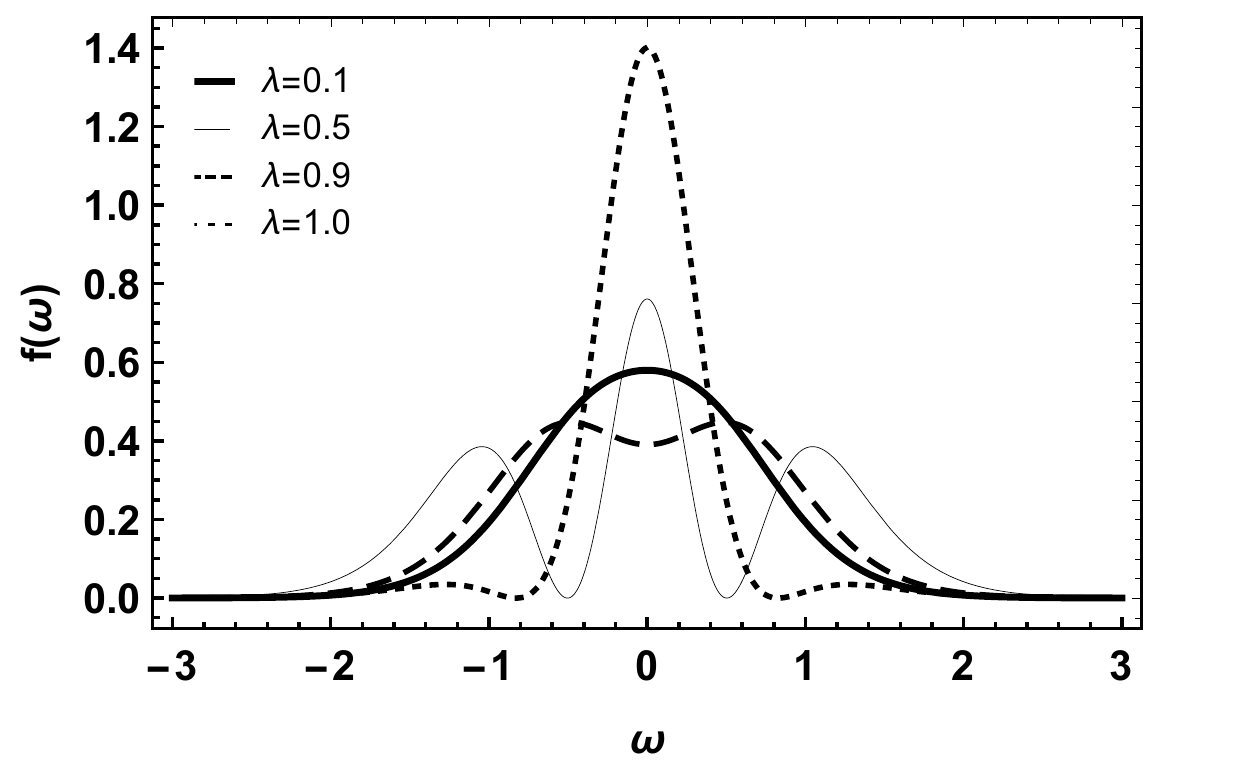}
\includegraphics[height=5.5cm,width=7cm]{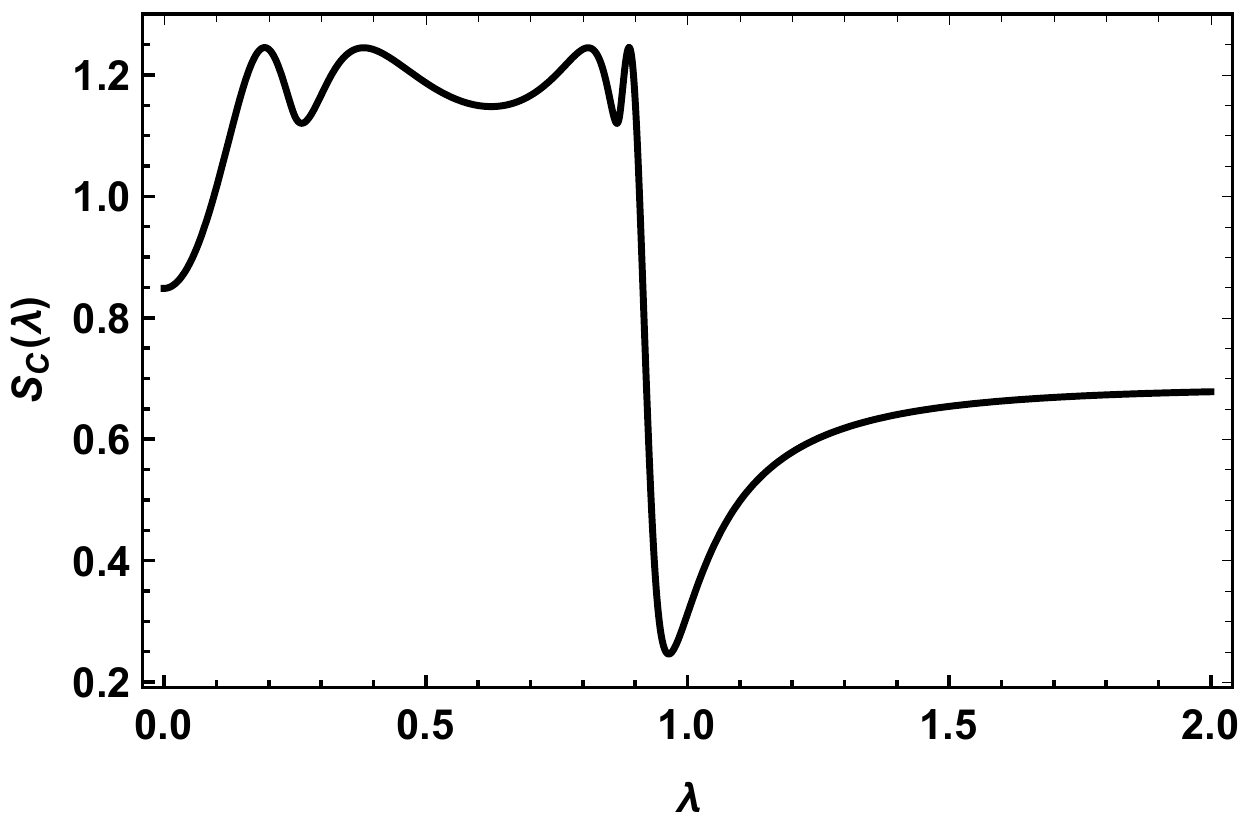}\\ 
(a) \hspace{7cm} (b)\\
\includegraphics[height=5.5cm,width=7cm]{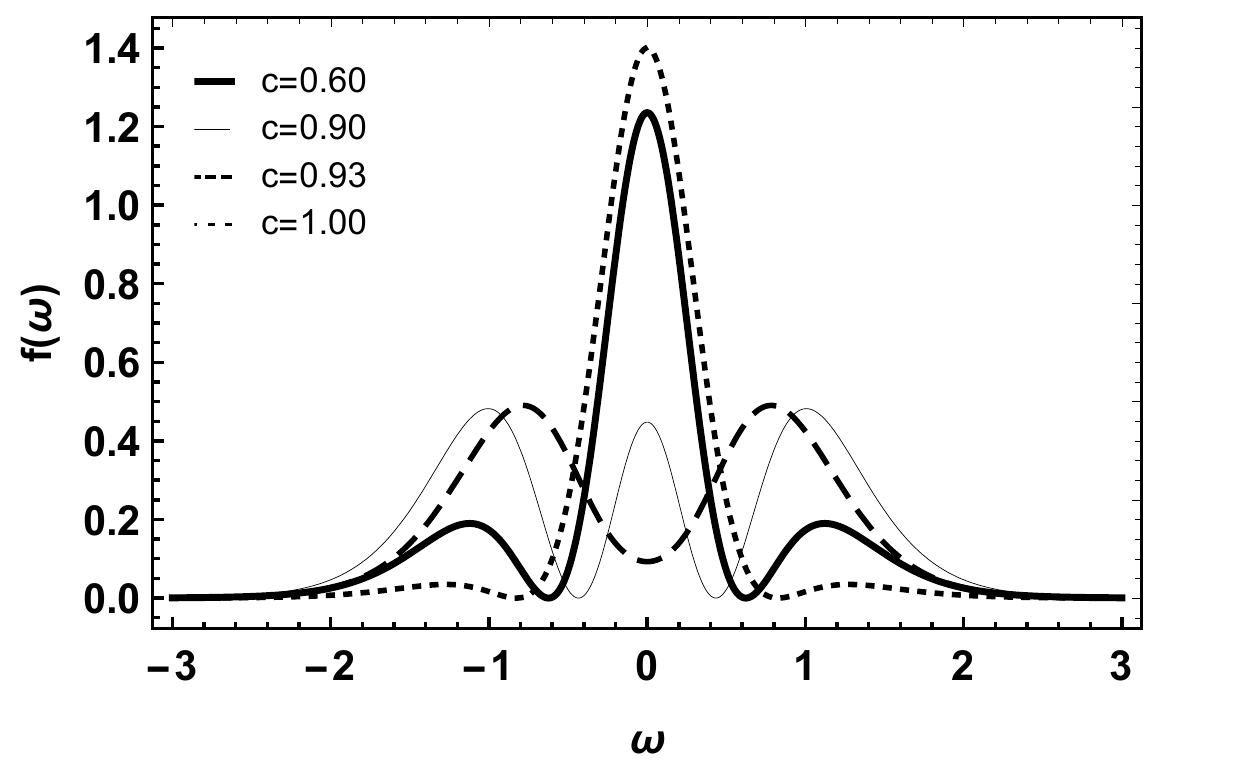}
\includegraphics[height=5.5cm,width=7cm]{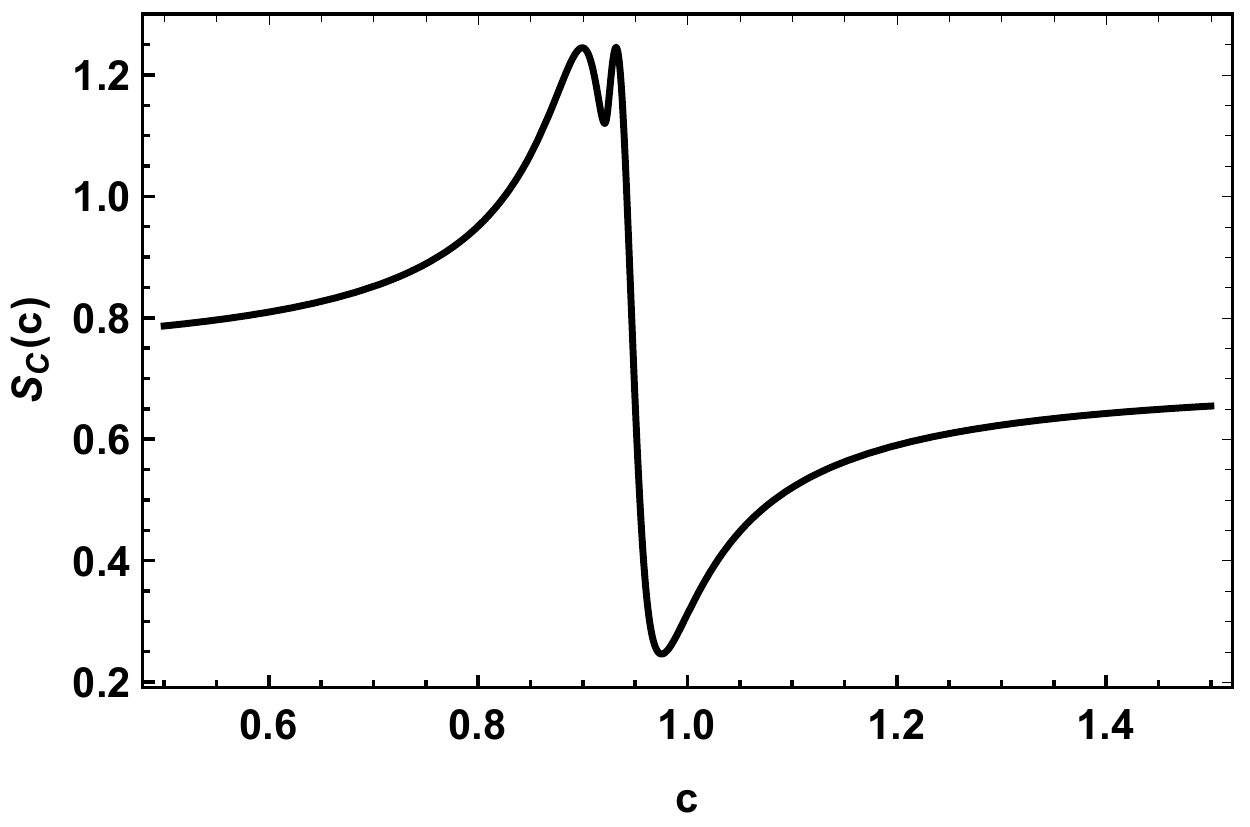}\\ 
(c) \hspace{7cm} (d)\\
\end{tabular}
\end{center}
\vspace{-0.5cm}
\caption{(a) Modal fraction of the $f_1(T)$ model with $n=1$, $k=-0.5$, and $c=1$.
(b) DCE with $n=1$, $k=-0.5$, and $c=1$. (c) Modal fraction of the $f_1(T)$ model with $n=1$, $k=-0.5$, and $\lambda=1$. (d) DCE with $n=1$, $k=-0.5$, and $\lambda=1$.
\label{fig922}}
\end{figure}

In this case, we noted that similarities between the results of the cases $A_2(y)$ and $A_1(y)$ appear. However, in the case $A_2(y)$, the most interesting results arise when the parameters $c$ and $\lambda$ are modified. The variation of these parameters in the model $f_1(T)$ with $n=1$ and $k=5$ leads us to an oscillatory profile of the modal fraction [see Figs. \ref{fig922}(a) and \ref{fig922}(c)]. The modal fraction profile is more localized when $c\approx 1$ and $\lambda\approx 1$, indicating the values of the parameters that describe the most likely field configurations. The DCE result [see Figs. \ref{fig922}(b) and \ref{fig922}(d)] confirms that the higher the value of the $c$ and $\lambda$ parameters, the lower the stability of the matter field, so the field configurations are less likely. Here, it is interesting to highlight the compact-like structures (more geometrically contracted structures) obtained in Sec. \ref{sec1} is the least likely configuration of the theory.

\section{Final remarks}
\label{sec3}

In this work, the solutions of the matter field were studied in braneworld with $f(T)$ teleparallel gravity. For this study, two models were considered, namely, the model $f_1(T)=T+kT^{n}$ and $f_{2}(T)=T+\alpha T^{2}+\beta T^ {3}$. To analyze these models it was necessary to particularize them through the ans\"{a}tz of the warp factor, i. e., $A_1(y)=-p\ln{\cosh(\lambda y)}$ and $A_2(y)=\ln\vert\tanh[\lambda(y+c)]-\tanh[\lambda(y-c)]\vert$. In both models, the parameters $k$, $n$, $\alpha$, and $\beta$ are responsible for the brane split and the deformation of the matter field. Furthermore, it was also possible to notice that the parameters $c$ and $\lambda$ of warp factor $A_2$ modify the brane solutions, so as obtain compact-like configurations.

DCE was an important tool that helped us by providing criteria to control the stability of our model's configurations based on informational content related to the brane. Indeed, this was possible because the DCE is proportional to the brane energy. Thus, using the concept of the DCE is possible to select the most likely (and stable) configurations to be found for our $f_{1,2}(T)$ models.

In the $A_1(y)$ case of the $f_1(T)$ model, it was possible to observe that the DCE has a minimum value when $k\approx -0.05$ ($n=2$), $0.005$ ($n=3 $), and $-0.004$ ($n=4$). Naturally, these results suggest the appearance of a second domain wall, implying the beginning of the deformation of the matter field and a double phase transition of the scalar field. This double phase transition appears reflected in the energy density that indicates a brane splitting. The same is true for the $f_2(T)$ model, but in this case, the minimum DCE configurations appear when $\alpha\approx-0.05$ and $\beta\approx0.005$.

For the warp factor $A_2(y)$, in the model $f_1(T)$, if $n=1$ and $k=-0.5$, it can be seen that the higher the value of $c$ and $\lambda$ smaller the stability of the structures. In other words, the higher the value of the $c$ and $\lambda$ parameters, the more compact the matter field profile will be. Therefore, compact-like (contracted) structures are less likely. Commonly, we found the most likely configuration when $c\approx1$ and $\lambda\approx1$.

Finally, the results found help us to have a better understanding of the topological structures (and its contraction). Indeed, DCE provides a complementary analysis of the matter field solutions, phase transitions, and brane splitting.

A future perspective of this study is to understand how our results are changed if the Lagrangian of the matter field has noncanonical dynamics \cite{LPA11}. Another possibility is to study the location of fermions and the massive modes of these models. We hope to carry out these studies soon.

\section*{Acknowledgments}
The authors thank the Conselho Nacional de Desenvolvimento Cient\'{i}fico e Tecnol\'{o}gico (CNPq), grants n$\textsuperscript{\underline{\scriptsize o}}$ 309553/2021-0 (CASA) and the Coordena\c{c}\~{a}o de Aperfei\c{c}oamento de Pessoal de N\'{i}vel Superior (CAPES), grants n$\textsuperscript{\underline{\scriptsize o}}$ 88887.372425/2019-00 (FCEL), for financial support.
%\bibliographystyle{lion-msc}
%\bibliography{refdis}

\end{document}